\documentclass[11pt]{article}
\usepackage{}
\usepackage{amssymb}
\usepackage{amsfonts}
\usepackage{mathrsfs}
\usepackage{graphicx}
\usepackage{epstopdf}
\usepackage{amsmath}
\usepackage{color}
\usepackage{amsthm}

\usepackage{pifont,bm}
\usepackage{tikz}
\usepackage{enumerate}
\newtheorem{thm}{Theorem}[section]

\newtheorem{prop}[thm]{Proposition}
\theoremstyle{definition}

\newtheorem{rem}[thm]{Remark}

\makeatletter
\def\@biblabel#1{[#1]}
\makeatother

\makeatletter \@addtoreset{equation}{section}

\begin{document}
%\begin{CJK*}{GBK}{song}

\begin{titlepage}
\title{\bf{Superregular solitonic solutions of the defocusing nonlocal nonlinear Schr\"{o}dinger equation and
their novel perspectives
\footnote{%This work was supported by the Natural Science Foundation of China under Grant No. 12201622.\protect\\
Corresponding authors.\protect\\
\hspace*{3ex} E-mail addresses: xbwang@cumt.edu.cn, xiubinwang@163.com (X.B. Wang)}
}}
\author{Duan-Yun Xia, Xiu-Bin Wang$^{*}$\\
%%%%%%%%%%%%%%%%%%%%%%%%%%%%%%%%%%%%%%%%%%%%%%%%%%%%%%%%%%%%%%%%%%%%%%%%%%%%%%%%%%%%%%%%%
%%%%%              以下两行为作者单位
%%%%%%%%%%%%%%%%%%%%%%%%%%%%%%%%%%%%%%%%%%%%%%%%%%%%%%%%%%%%%%%%%%%%%%%%%%%%%%%%%%%%%%%%%
\small \emph{School of Mathematics, China University of Mining and Technology, Xuzhou 221116, P.R. China}
%\small \emph{}\\
\date{}}
\thispagestyle{empty}
\end{titlepage}
\maketitle

\vspace{-0.5cm}
\begin{center}
\rule{15cm}{1pt}\vspace{0.3cm}

\parbox{15cm}{\small
{\bf Abstract}\\
\hspace{0.5cm} Previous studies have indicated that the local defocusing nonlinear Schr\"{o}dinger equation (NLSE) is modulationally stable.
Therefore, superregular solitonic solutions characterizing the nonlinear evolution of modulational instability under weak localized perturbations do not exist.
In this work, we study the nonlinear stage of the modulation instability of a condensate in the framework of the
defocusing nonlocal NLSE.
By using the dressing method and Joukowski transformation, we systematically derive the general $N$-solitonic solution of the
defocusing nonlocal NLSE, and the explicit expressions for one-solitonic, two-solitonic and superregular two-solitonic solutions are provided.
Then dynamics of these solutions is also examined.
We find that these solutions show a much wider variety than those in the local NLSE, such as the one-solitonic solution composed of two interacting wave components and the two-solitonic solution containing four wave components.
Finally, the accuracy and noise robustness of the analytical solutions are confirmed by numerical simulations.
Such results advance the theory of superregular solitonic solutions in nonlocal systems and offer new physical understanding of wave propagation in nonlocal nonlinear media.
}

\vspace{0.5cm}
\parbox{15cm}{\small{

\vspace{0.3cm} \emph{Key words:} The reverse-time nonlocal nonlinear Schr\"{o}dinger equation; Dressing method; Superregular solitonic solutions.

\emph{PACS numbers:}  02.30.Ik, 05.45.Yv, 04.20.Jb. } }
\end{center}
\vspace{0.3cm} \rule{15cm}{1pt} \vspace{0.2cm}

\section{Introduction}
Breathers are important localized periodic solutions to nonlinear integrable equations and give an exact analytical description of the nonlinear saturation of modulation instability (MI) \cite{ref1,ref2,ref3,ref4,ref5}. In contrast to solitons, breathers possess spatially localized envelopes with high-frequency oscillations inside. This "breathing" behavior makes them a fundamental model for investigating energy localization and periodic dynamics.
Physically, breathers arise in a variety of physical systems including optical fiber communications \cite{ref6}, water wave dynamics \cite{ref7},
and Bose-Einstein condensates \cite{ref8}. Generally speaking, breathers fall into two categories: the Akhmediev breather and the Kuznetsov-Ma soliton. They describe distinct dynamical mechanisms: the former propagates periodically in space and is localized in time \cite{ref2,ref9,ref10}, whereas the latter is spatially localized and exhibits periodic breathing in time \cite{ref1,ref11}.
When the spatial or temporal period is taken to be infinite, both reduce to Peregrine soliton in the
limit, which provides an important theoretical model for revealing the physical mechanism of rogue wave excitation induced by MI \cite{ref12,ref13,ref14}.
%Both can reduce to the Peregrine soliton \cite{ref13} under specific parameter limits,

However, traditional breather theory has a fundamental limitation. Both Akhmediev breather and Kuznetsov-Ma soliton develop MI starting from periodic or highly regular perturbations \cite{ref15,ref16}. In reality, localized small perturbations are more common, rather than infinitely extended periodic modulations \cite{ref17}. This gap was not filled until Zakharov and Gelash proposed the concept of superregular breathers in 2013 \cite{ref18}. The superregular breathers describe how a small localized perturbation evolves under MI into a pair of quasi-Akhmediev breathers. They travel in opposite directions with a small angle and nearly annihilate to recover the initial small perturbation upon collision \cite{ref19}. Superregular breathers carry important physical meaning. They can be regarded as high-order exact solutions of the scalar nonlinear Schr\"{o}dinger equation (NLSE) \cite{ref16,ref20,ref21}. The explicit expression of this solution can be derived by the dressing method combined with the Joukowski transformation. It extends the nonlinear stage of MI from a special case of periodic perturbations to a general solution for localized perturbations. It also builds a bridge between linear stability analysis and full nonlinear evolution via the exact correspondence between group velocity and growth rate \cite{ref22}. Since it was proposed, superregular breathers have been extended from the scalar NLSE to the derivative NLSE \cite{ref20}, complex modified KdV equation \cite{ref23}, Hirota equation \cite{ref24}, coupled Maxwell-Bloch equation \cite{ref25} and vector models with high-order effects \cite{ref15,ref26,ref27}. They have also been verified in both optical and hydrodynamic experiments \cite{ref28}. Notably, the characteristic restoration of the background after such nonlinear interactions is deeply connected to the Fermi-Pasta-Ulam (FPU) recurrence. This phenomenon traces its historical roots to the dislocation dynamics on periodic lattices in the Korteweg-de Vries (KdV) equation \cite{KEA1}, and has been subsequently explained for the NLSE via the soliton lattice concept \cite{KEA2}, a mechanism recently reviewed and confirmed by experimental observations \cite{KEA3}.

Although studies on superregular breathers are well developed for local systems, their extension to nonlocal frameworks faces unique structural challenges. In 2013, Ablowitz and Musslimani applied $\mathcal{PT}$-symmetry constraints to the AKNS system and first put forward the nonlocal NLSE \cite{ref29}
\begin{equation*}
\textrm{i}\psi_t(x,t)+\frac{1}{2}\psi_{xx}(x,t)+\sigma\psi^2(x,t)\psi^*(-x,t)=0.
\end{equation*}
Here, $\sigma=\pm1$ is the sign of nonlinearity, where \(\sigma=1\) corresponds to the focusing regime and \(\sigma=-1\) corresponds to the defocusing regime, and the asterisk $*$ represents complex conjugation. The nonlocal term $\psi^*(-x,t)$ in this equation breaks the traditional framework of local integrable equations. This breakthrough motivates systematic studies on nonlocal integrable systems. At the equation level, various nonlocal equations, such as reverse space-time \cite{ref30} and nonlocal derivative equation \cite{ref31}, were constructed. At the solution level, exact solutions such as solitonic solutions and breathers for these equations have also been systematically established \cite{ZJJJ1,ZJJJ2,ZJJJ3,ZJJJ4,ZJJJ5,ZJJJ6,ref32,ref33,ref34,XBWWW1,ref35,ref36}, forming a new branch of integrable system theory. This revolutionary extension greatly enriches the family of integrable equations. It also produces many distinctive dynamical effects, such as unique interaction patterns between nonlocal solitons \cite{ref37}, abnormal transport of energy and momentum under nonlocal coupling \cite{ref38,ref39}, and phase transitions related to symmetry breaking and recovery in dynamical processes \cite{ref40}. As another nonlocal reduction of the AKNS system, the reverse-time nonlocal NLSE couples the field in the nonlinear term with its time-reversed counterpart. It introduces extra symmetry constraints in the spectral problem and changes the construction of solitonic solutions as well as the criterion of MI. Its explicit expression reads
\begin{equation}\label{eq1}
\textrm{i}\psi_t(x,t)+\frac{1}{2}\psi_{xx}(x,t)+\sigma\psi^2(x,t)\psi(x,-t)=0.
\end{equation}
Existing studies have constructed standard solitons, nondegenerate multi-solitons \cite{ref41} and breathers of this equation via the Darboux transformation \cite{ref42,ref43,ref44} and Riemann-Hilbert method \cite{ref45,ref46}. However, to our knowledge, superregular solitonic solutions have not yet been systematically investigated under the reverse-time nonlocal framework.

As reported in previous investigations, MI is absent for the NLSEs in the defocusing regime.
Therefore, superregular solitonic solutions that characterize the nonlinear stage of MI do not exist.
Recently, higher-order rogue wave solutions of the reverse-time nonlocal NLSE in the defocusing regime have been successfully constructed in \cite{ref44}.
This raises questions worthy of exploration:

$\bullet$ Can superregular solitonic solutions be constructed for Eq.\eqref{eq1} in the defocusing regime?

$\bullet$ What properties do such superregular solitonic solutions possess in the defocusing regime?

$\bullet$ What will happen in the nonlinear stage of MI with the time-reversed counterpart for Eq.\eqref{eq1} in the defocusing regime?

This work aims to address the above issues by investigating the defocusing regime of Eq.\eqref{eq1} via the dressing method and Joukowski transformation. First, the $N$-solitonic and superregular solitonic solutions of the equation are analytically derived. Furthermore, dynamical characteristics of these solutions are explored through parametric adjustment and graphical analysis, revealing unique multi-wave coupling interactions.
These patterns differ from the standard NLSE examples considered here. Our results reveal the characteristics of the
nonlinear stage of MI induced by the reverse-time nonlocal framework.

%The main contributions of this work are twofold.
%\begin{itemize}
%\item In the defocusing regime, superregular solitonic solutions of the reverse-time nonlocal NLSE are systematically derived via the dressing method combined with the Joukowski transformation, with particular emphasis on the construction of superregular two-solitonic solutions. This fills the gap in the study of superregular breathers for reverse-time nonlocal systems.
%\item Dynamical behaviors of the solutions distinct from those of the standard NLSE are revealed: the one-solitonic solution is formed by two interacting wave components, the two-solitonic solution contains four wave components, and so on. This reflects the profound influence of the reverse-time nonlocal term on the solution structure.
%\end{itemize}

The paper is organized as follows. In Section 2, the spectral problem in the defocusing regime is simplified, and the general formula for $N$-solitonic solution is derived by the dressing method. Section 3 presents a detailed analysis of the explicit expressions and dynamical behaviors of one-solitonic and two-solitonic solutions in the defocusing regime. Sections 4 and 5 are devoted to soliton annihilation and superregular two-solitonic solution for the defocusing case, respectively. Numerical simulations are employed to verify the accuracy and robustness of the solutions in Section 6. Finally, Section 7 draws the conclusions.

\section{General $N$-solitonic solution}\label{sec2}

For Eq.\eqref{eq1} in the defocusing regime, direct spectral analysis is intractable. Instead, we consider the linear system associated with the transformed field $\varphi$ from gauge transformation $\psi=\textrm{i}e^{\textrm{i}t}\varphi$, which reads for a $2\times2$ matrix function $\Phi$
\begin{align}
&\Phi_{x}=U\Phi,~~ U=\lambda\sigma_3+Q,\label{eq3}\\
&\Phi_{t}=V\Phi,~~ V=\textrm{i}\lambda (\lambda\sigma_3+Q)+\frac{\textrm{i}}{2}\sigma_3(Q_x-Q^2-\textbf{I}),\label{eq4}
\end{align}
where
\begin{equation}
\sigma_3=\begin{pmatrix}
1 & 0\\
0 & -1
\end{pmatrix},\quad
Q=\begin{pmatrix}
0 & \varphi(x,t)\\
-\varphi(x,-t) & 0
\end{pmatrix},\nonumber\\
\label{matrix_uw}
\end{equation}
and $\textbf{I}$ represents the $2\times2$ identity matrix.

\begin{prop}\label{prop:1}
Introduce a gauge transformation of the form $\psi = D \varphi$, where $D$ denotes a given invertible transformation operator. If $\varphi$ solves the transformed evolution equation, then $\psi$ is a solution to the original governing equation.
\end{prop}

Guided by the above proposition, we first derive analytical solutions for $\varphi$ of the simplified model, and reconstruct wavefunctions satisfying the original equation by substituting $\psi = \textrm{i}e^{\textrm{i}t}\varphi$.
From equations \eqref{eq3} and \eqref{eq4}, we derive an auxiliary system for $\Phi^{-1}$ and the corresponding adjoint system for $\Phi^\top$:
\begin{align*}
\Phi^{-1}_x &= -\Phi^{-1}U,\quad \Phi^{-1}_t = -\Phi^{-1}V,\\
\Phi^\top_x &= \Phi^\top U^\top,\quad \Phi^\top_t = \Phi^\top V^\top.
\end{align*}
By substituting $\lambda\to-\lambda$ and $t\to-t$ into the adjoint system, we obtain the matrix identities
\begin{equation*}
U^\top(-\lambda,-t)=- U(\lambda,t),\quad V^\top(-\lambda,-t)= V(\lambda,t).
\end{equation*}
This implies that the two systems are identical, so there exists a family of solutions $\Phi$ satisfying the constraint
\begin{equation}\label{key_constraint}
\Phi^\top(-\lambda,-t)=\Phi^{-1}(\lambda,t).
\end{equation}

In the following analysis, we assume that this constraint holds as a fundamental condition. The core principle behind the dressing technique is outlined below. Assume a known particular solution $\varphi_0$ for the transformed equation, alongside its fundamental matrix-valued solution $\Phi_0$ defined over the variables $x,t,\lambda$. This fundamental matrix obeys the associated coupled linear differential system:
\begin{align*}
\Phi_{0x} = U_0\Phi_0, \quad \Phi_{0t} = V_0\Phi_0.
\end{align*}
Here the coefficient matrices $U_0$ and $V_0$ originate from substituting $\varphi \to \varphi_0$ into the $Q$.

Next we introduce the core transformation kernel termed the dressing operator via the definition:
\begin{equation*}
\chi = \Phi\Phi_0^{-1}.
\end{equation*}
It should be noted that the matrix $\Phi$ remains an undetermined quantity at this stage. We impose the analytic constraint that $\chi$ is normalized in the asymptotic limit $|\lambda|\to\infty$, with the large-$\lambda$ expansion
\begin{equation}\label{eq7}
\chi(\lambda) \to E + \frac{N}{\lambda} + O(\lambda^{-2}),\quad \text{at } |\lambda|\to\infty,
\end{equation}
where $N$ denotes a $\lambda$-independent matrix coefficient $N(x,t)$ and $E$ stands for the identity matrix. Making use of relation \eqref{key_constraint}, we can further derive a reduction condition for $\chi(x,t,\lambda)$:
\begin{equation}\label{eq8}
\chi^T(-\lambda,-t) = \chi^{-1}(\lambda,t).
\end{equation}
The dressing function $\chi$ is subject to an overdetermined set of linear equations, the first of which takes the form
\begin{equation}\label{eq9}
\chi_x = U\chi - \chi U_0,
\end{equation}
while its inverse counterpart satisfies
\begin{equation}\label{eq10}
\chi_x^{-1} = -\chi^{-1}U + U_0\chi^{-1}.
\end{equation}
Eq.\eqref{eq10} can be rewritten as follows
\begin{equation}\label{eq11}
U = -\chi\left(\frac{\partial}{\partial x}-U_0\right)\chi^{-1}.
\end{equation}
Similarly, the following relation can be derived from Eq.\eqref{eq4}:
\begin{equation}\label{eq12}
V = -\chi\left(\frac{\partial}{\partial t}-V_0\right)\chi^{-1}.
\end{equation}

Once we fix $\chi$ to make the expression of $U$ in Eq.\eqref{eq11} and $V$ in Eq.\eqref{eq12} nonsingular everywhere on the $\lambda$-plane, a novel exact solution of the governing Eq.\eqref{eq3} and \eqref{eq4} can be generated. Substituting the asymptotic expansion \eqref{eq7} into differential relation \eqref{eq9} yields the explicit algebraic transformation known as the dressing recursion relation:
\begin{equation*}
\varphi = \varphi_0 - 2N_{12}.
\end{equation*}
From Proposition \ref{prop:1}, we obtain the following formula:
\begin{equation}\label{eq13}
\psi = \textrm{i}(\varphi_0 - 2N_{12})e^{\textrm{i}t}.
\end{equation}

Next, we derive the dressing formula in the case where there are singularities in $U$ and $V$.
$\Phi_0(x,t,\lambda)$ is assumed to be known, while $\chi$ denotes a rational function of the spectral parameter $\lambda$ given by
\begin{equation*}
\chi = E + \sum_{m} \frac{N_m}{\lambda - \lambda_m}.
\end{equation*}
Without loss of generality, we may take $\operatorname{Re}(\lambda_n) > 0$. Owing to the symmetry condition \eqref{eq8} satisfied by $\chi$, its inverse admits the representation
\begin{equation*}
\chi^{-1} = E - \sum_{m} \frac{ N_m^T(-\lambda,-t)}{\lambda + \lambda_m}.
\end{equation*}
Let us denote
\begin{equation}\label{eq16}
\chi_n := \left. \chi \right|_{\lambda = -\lambda_n} = E - \sum_{m} \frac{N_m}{\lambda_n + \lambda_m},
\end{equation}
and consider the fundamental identity
\begin{equation}\label{eq17}
\chi \chi^{-1} = E.
\end{equation}
Substituting $\lambda = -\lambda_n$ into the fundamental identity \eqref{eq17} gives
\begin{equation}\label{eq18}
\chi_n N_n^T(-\lambda_n,-t) = 0.
\end{equation}
It follows immediately that $N_n$ and $N_n^T$ are degenerate matrices, as are $\chi_n$ and $\chi_n^T$. We introduce two sets of complex-valued vectors $\{\mathbf{p}_n\}$ and $\{\mathbf{q}_n\}$. Each vector has two components, i.e., the indices $\alpha,\beta$ run over $\{1,2\}$, and we set
\begin{equation}\label{eq19}
N_{n,\alpha\beta} = p_{n,\alpha}q_{n,\beta}, \qquad N_{n,\alpha\beta}^T = q_{n,\alpha}p_{n,\beta},
\end{equation}
where $n$ labels the $n$-th vector.
Substituting \eqref{eq19} into condition \eqref{eq18}, we arrive at the vector equation
\begin{equation}\label{eq20}
\chi_n q_{n,\alpha}(-\lambda_n,-t) = 0.
\end{equation}
The matrix elements of $\chi$ and its inverse take the form
\begin{align}\label{eq21}
\chi_{\alpha\beta} &= \delta_{\alpha\beta} + \sum_{n} \frac{p_{n,\alpha}q_{n,\beta}}{\lambda - \lambda_n},\nonumber \\
\chi_{\alpha\beta}^{-1} &= \delta_{\alpha\beta} - \sum_{n} \frac{ q_{n,\alpha}(-\lambda,-t)p_{n,\beta}(-\lambda,-t)}{\lambda + \lambda_n}.
\end{align}
Inserting \eqref{eq21} into \eqref{eq11} produces a function U with poles at \(\lambda = \lambda_n\) and \(\lambda = -\lambda_n\) in the general case. For the dressing procedure to be valid, residues at all poles must cancel. Analysis of Eq.\eqref{eq11} in a neighborhood of $\lambda = -\lambda_n$ shows that cancellation of the residue at this point requires
\begin{equation*}
\frac{\partial \mathbf{q}_n(-t)}{\partial x} = U_0(-\lambda_n) \mathbf{q}_n(-t).
\end{equation*}
This equation admits the explicit solution
\begin{equation}\label{eq23}
\mathbf{q}_n(-t) = \Phi_0(-\lambda_n) \zeta_n.
\end{equation}
Here, $\zeta_n$ is an arbitrary complex-valued vector. Subsequent notation is defined as
\begin{equation}\label{eq24}
F_n = \Phi_0(-\lambda_n).
\end{equation}
From this moment, the set of vectors $\{\mathbf{q}_n\}$ is taken as known. Finding the second set $\{\mathbf{p}_n\}$ requires solving relation \eqref{eq20}, which is equivalent to a system of linear equations. Substituting \eqref{eq16} and \eqref{eq19} into \eqref{eq20} gives
\begin{equation}\label{eq25}
 \mathbf{q}_n = \sum_{m} \frac{\mathbf{q}_m^T(-t) \mathbf{q}_n}{\lambda_n + \lambda_m} \mathbf{p}_m(-t).
\end{equation}
Let
\begin{equation}\label{eq26}
M_{nm} = \frac{\mathbf{q}_m^T(-t) \mathbf{q}_n}{\lambda_n + \lambda_m},
\end{equation}
and let $M = \det(M_{nm})$. Then equation \eqref{eq25} can be written as
\begin{equation*}
 \mathbf{q}_n(-t) = \sum_{m} M_{nm}(-t) \mathbf{p}_m.
\end{equation*}
The matrix $N$ comes from the asymptotic expansion of $\chi$ \eqref{eq7}, which reads
\begin{equation*}
N_{\alpha\beta} = \sum_{n} p_{n,\alpha} q_{n,\beta}.
\end{equation*}
This sum can be evaluated as a ratio of determinants:
\begin{equation*}
N_{\alpha\beta} = -\frac{\widetilde{M}_{\alpha\beta}}{M}.
\end{equation*}
Here, $\widetilde{M}_{\alpha\beta}$ denotes the determinant
\begin{equation}\label{eq30}
\widetilde{M}_{\alpha\beta} =
\begin{vmatrix}
\delta_{\alpha\beta} & q_{1,\beta} & \cdots & q_{n,\beta} \\
q_{1,\alpha}(-t) & & & \\
\vdots & & M_{nm}(-t) & \\
q_{n,\alpha}(-t) & & &
\end{vmatrix}.
\end{equation}
Eq.\eqref{eq13} can then be written as
\begin{equation*}
\varphi =\varphi_0 + 2\frac{\widetilde{M}_{12}}{M}.
\end{equation*}
Thus,
\begin{equation}\label{eq31}
\psi=\textrm{i}\left(\varphi_0 + 2\frac{\widetilde{M}_{12}}{M}\right)e^{\textrm{i}t}.
\end{equation}

The initial solution is taken as $\psi_0=\textrm{i}e^{\textrm{i}t}$, namely $\varphi_0=1$. The solution exhibits instability under small perturbations. The corresponding instability growth rate reads
\begin{equation*}
g(\kappa)=|\kappa|\sqrt{1-\kappa^2/4},
\end{equation*}
where $\kappa$ represents the spatial wavenumber of the perturbation. Exponential growth occurs in the range $0<|\kappa|<2$, with the maximum growth rate attained at $|\kappa|=\sqrt{2}$. Under this condition, $U_0$ and $V_0$ are respectively:
\begin{equation*}
U_0 = \begin{pmatrix}
\lambda & 1 \\
-1 & -\lambda
\end{pmatrix}, \quad
V_0=\begin{pmatrix}
\textrm{i}\lambda^2 & \textrm{i}\lambda \\
-\textrm{i}\lambda & -\textrm{i}\lambda^2
\end{pmatrix}.
\end{equation*}
Then the fundamental solution $\Phi_0$ to Eq.\eqref{eq3} and \eqref{eq4} is given by
\begin{equation*}
\Phi_0(x,t,\lambda) = \frac{1}{\sqrt{1 - s^2(\lambda)}}
\begin{pmatrix}
\exp(\phi(x,t,\lambda)) & s(\lambda)\exp(-\phi(x,t,\lambda)) \\
s(\lambda)\exp(\phi(x,t,\lambda)) & \exp(-\phi(x,t,\lambda))
\end{pmatrix}.
\end{equation*}
Here
\begin{equation*}
\phi = kx + \Omega t, \quad k^2 = \lambda^2 - 1,\quad
\Omega = \textrm{i}\lambda k, \quad s(\lambda) = -\frac{1}{\lambda + k}.
\end{equation*}
Then
\begin{equation*}
\Phi_0^{-1}(x,t,\lambda) = \frac{1}{\sqrt{1 - s^2(\lambda)}}
\begin{pmatrix}
\exp(-\phi(x,t,\lambda)) & -s(\lambda)\exp(-\phi(x,t,\lambda)) \\
-s(\lambda)\exp(\phi(x,t,\lambda)) & \exp(\phi(x,t,\lambda))
\end{pmatrix}.
\end{equation*}
Take the branch of $k(\lambda) = \sqrt{\lambda^2 - 1}$ on the complex plane such that $k(\lambda) \sim \lambda$ as $|\lambda| \to \infty$ and $k(-\lambda) = -k(\lambda)$. Then
\begin{equation}\label{eq32}
s(-\lambda)=-s(\lambda), \quad \phi(x,-t,-\lambda)=-\phi(x,t,\lambda).
\end{equation}
Define $\phi_n$ and $s_n$ as follows:
\begin{equation*}
s_n=s(\lambda_n), \quad \phi_n=\phi(\lambda_n).
\end{equation*}
From Eq.\eqref{eq32}, we have
\begin{equation*}
s(-\lambda_n)= -s_n, \quad \phi(-t,-\lambda_n)= -\phi_n.
\end{equation*}
Then
\begin{equation}\label{eq33}
F_n = \Phi_0(-\lambda_n) =
\begin{pmatrix}
\exp(-\phi_n(-t)) & -s_n \exp(\phi_n(-t)) \\
-s_n \exp(-\phi_n(-t)) & \exp(\phi_n(-t))
\end{pmatrix}.
\end{equation}
The coefficient $(\sqrt{1 - s^2(\lambda)})^{-1}$ is omitted, since it does not influence the final result. Set $\zeta_n = \begin{pmatrix} 1 \\ C_n \end{pmatrix}$. Then, from Eqs.\eqref{eq23}, \eqref{eq24}, and \eqref{eq33}, it follows that
\begin{equation*}
q_{n1} = \exp(-\phi_n) - C_n s_n \exp(\phi_n), \quad
q_{n2} = - s_n \exp(-\phi_n) + C_n \exp(\phi_n).
\end{equation*}
Now we use the Joukowsky transform and define the variables as follows:
\begin{equation*}
\lambda_n = \frac{1}{2}\left(\xi_n +\xi_n^{-1}\right), \quad
k_n = \frac{1}{2}\left(\xi_n -\xi_n^{-1}\right), \quad
s_n = -\xi_n^{-1}.
\end{equation*}
With the above variables
\begin{equation*}
F_n =
\begin{pmatrix}
\exp(-\phi_n(-t)) & \xi_n^{-1} \exp(\phi_n(-t)) \\
\xi_n^{-1} \exp(-\phi_n(-t)) & \exp(\phi_n(-t))
\end{pmatrix},
\end{equation*}
and
\begin{equation*}
q_{n1} = \exp(-\phi_n) + C_n \xi_n^{-1} \exp(\phi_n), \quad
q_{n2} = \xi_n^{-1} \exp(-\phi_n) + C_n \exp(\phi_n).
\end{equation*}
We now introduce the parametrization
\begin{equation*}
R_n = \exp(z_n),\quad \xi_n = R_n \exp(\textrm{i}\alpha_n),\quad C_n = \exp(\textrm{i}\theta_n + \mu_n),\quad w_n = -\textrm{i}\alpha_n - z_n.
\end{equation*}
Thus, the parametrized expressions for $\lambda_n$ and $k_n$ are given by:
\begin{align*}
\lambda_n=\cos\alpha_n \cosh z_n + \textrm{i} \sin\alpha_n \sinh z_n,~~
k_n=\cos\alpha_n\sinh z_n+\textrm{i}\sin\alpha_n\cosh z_n.
\end{align*}
By substituting $\lambda_n$ and $k_n$ into $\phi_n$ and performing a redefinition, the expression for $q_n$ is derived as follows:
\begin{equation}\label{qn}
q_{n1} = \exp(-\phi_n) + \exp(w_n + \phi_n),\quad  q_{n2} = \exp(w_n - \phi_n) + \exp(\phi_n).
\end{equation}
Here
\begin{align*}
\phi_n &= u_n + \textrm{i} v_n, \\
u_n &= \text{\ae}_n x + \gamma_n t + \frac{\mu_n}{2}, & v_n &= \beta_n x + \omega_n t + \frac{\theta_n}{2}, \\
\text{\ae}_n &= \cos\alpha_n \sinh z_n, & \gamma_n &= -\frac{1}{2} \sin 2\alpha_n \cosh 2z_n, \\
\beta_n &= \sin\alpha_n \cosh z_n, & \omega_n &= \frac{1}{2} \cos 2\alpha_n \sinh 2z_n.
\end{align*}
Substitution of Eq.\eqref{qn} into Eq.\eqref{eq31} yields the $N$-solitonic solution under the defocusing regime.

The $N$-solitonic solution possesses invariance under space translations. Substitute $x \to x-x_0$, then each coefficient $C_n$ in the solution is multiplied by an exponential factor $\exp(2(\text{\ae}_n x_0)+2\textrm{i}(\beta_n x_0)$. This indicates that the position of the soliton can be arbitrarily shifted, while its essential physical nature remains unchanged. The $N$-solitonic solution corresponds to the nonlinear superposition of $N$ separate solitons. The group and phase velocities of an individual soliton are given by
\begin{equation*}
V_{gr}=\frac{\gamma_n}{\text{\ae}_n}=-\frac{\sin\alpha_n\cosh2z_n}{\sinh z_n},~~
V_{ph}=\frac{\omega_n}{\beta_n}=\frac{\cos2\alpha_n\sinh z_n}{\sin\alpha_n }.
\end{equation*}
If all group velocities are different and $t\to\pm\infty$, the $N$-solitonic solution slowly splits into separate solitons that move far apart. The asymptotic properties of the $N$-solitonic solution as $x\to\pm\infty$ are as follows
\begin{equation*}
\psi\rightarrow -\textrm{i}\exp(\textrm{i}t+\textrm{i}\alpha^\pm),~~x\rightarrow \pm\infty.
\end{equation*}
For widely separated multi-solitons, the total phase shift originates from the mutual interactions during soliton collisions, and the phase shift takes the following form
\begin{equation*}
\alpha^+=2(\alpha_1+\cdots+\alpha_N),~~\alpha^-=-2(\alpha_1+\cdots+\alpha_N).
\end{equation*}

For the reverse-time nonlocal NLSE Eq.\eqref{eq1}, a regular solitonic solution must satisfy two conditions: the single-valuedness of the asymptotic background and the absence of zeros in the denominator. The single-valuedness of the background requires that the two asymptotic phases are equal, i.e., $\alpha^+=\alpha^-$, which yields $\sum\alpha_n=0,\pm\pi/2$. It can be proven that the denominator is always nonvanishing when $\sum\alpha_n=0$ and all poles are distinct. In contrast, for $\sum\alpha_n=\pm\pi/2$, zeros may appear in the denominator under certain parameters, leading to singularities of the solution.
Therefore, the condition for the $N$-solitonic solution to be regular is
\begin{equation*}
\alpha_1+\cdots+\alpha_N=0.
\end{equation*}

\section{One-solitonic and two-solitonic solutions}\label{sec3}

This section is devoted to constructing the one-solitonic and two-solitonic solutions in the defocusing regime.
\subsection{One-solitonic solution}
To proceed with the derivation, we assume that the function $\chi$ has a single pole at $\lambda$. Correspondingly, its inverse $\chi^{-1}$ possesses a pole at $-\lambda$. From Eq.\eqref{eq25}, we obtain the relation between vectors {$\mathbf{p}$ and $\mathbf{q}$}:
\begin{equation*}
p_{\alpha} = \frac{2\lambda q_{\alpha}(-t)}{q_{1}q_{1}(-t) +  q_{2}q_{2}(-t)}.
\end{equation*}
Omitting the subscript $n$ in Eq.\eqref{eq19} gives
\begin{equation*}
N_{\alpha\beta}=\frac{2\lambda q_{\alpha}(-t)q_{\beta}}{q_{1}q_{1}(-t) + q_{2}q_{2}(-t)}.
\end{equation*}
Substitution into Eq.\eqref{eq13} yields
\begin{equation}\label{eq34}
\psi=\textrm{i}\left(\varphi_0 - \frac{4\lambda q_{1}(-t)q_{2}}{q_{1}q_{1}(-t) +  q_{2}q_{2}(-t)}\right)e^{\textrm{i}t}.
\end{equation}
The expressions of $q_1, q_2$ are similar to the counterparts derived in Eq.\eqref{qn}, given by
\begin{equation}\label{eq35}
q_{1} = \exp(-\phi) + \exp(w + \phi), \quad q_{2} = \exp(w - \phi) + \exp(\phi).
\end{equation}
Here
\begin{align*}
\phi &= u + \textrm{i} v, \\
u &= \text{\ae} x + \gamma t + \frac{\mu}{2}, & v &= \beta x + \omega t + \frac{\theta}{2}, \\
\text{\ae} &= \cos\alpha \sinh z, & \gamma &= -\frac{1}{2} \sin 2\alpha \cosh 2z, \\
\beta &= \sin\alpha \cosh z, & \omega &= \frac{1}{2} \cos 2\alpha \sinh 2z.
\end{align*}
The one-solitonic solution is governed by four parameters $z$, $\alpha$, $\mu$ and $\theta$.

Substitute Eq.\eqref{eq35} into Eq.\eqref{eq34} and set $\mu=0$, $\theta=\pi$, the explicit expression for the one-solitonic solution reads
\begin{equation}\label{eq36}
\psi=\textrm{i}\left(-1-\frac{A}{B}\right)e^{\textrm{i}t}.
\end{equation}
Here
\begin{align*}
A &= \cos(2\alpha + 2\omega t)\cosh(2z + 2\gamma t) - \cos(2\omega t)\cosh(2\gamma t) \nonumber\\
&\quad + \textrm{i}\sin(2\alpha + 2\omega t)\sinh(2z + 2\gamma t) - \textrm{i}\sin(2\omega t)\sinh(2\gamma t), \nonumber \\
B &= -\cos\alpha\cosh z\cos(2\beta x)\cosh(2\text{\ae} x) + \sin\alpha\sinh z\sin(2\beta x)\sinh(2\text{\ae} x) \nonumber\\
&\quad + \cos(2\omega t)\cosh(2\gamma t) - \textrm{i}\cos\alpha\cosh z\sin(2\beta x)\sinh(2\text{\ae} x) \nonumber\\
&\quad - \textrm{i}\sin\alpha\sinh z\cos(2\beta x)\cosh(2\text{\ae} x) + \textrm{i}\sin(2\omega t)\sinh(2\gamma t).
\end{align*}

Setting $\alpha=0$ in Eq.\eqref{eq36} reduces the solution to
\begin{equation}
\psi= \textrm{i}\left(\frac{\cosh z \cosh(2\text{\ae} x)-\cos(2\omega t)\cosh2z -\textrm{i}\sin(2\omega t)\sinh2z }{-\cosh z \cosh(2\text{\ae} x)+\cos(2\omega t)}\right)e^{\textrm{i}t}.
\end{equation}
Here
\begin{equation*}
\text{\ae}=\sinh z, \quad \omega=\frac{1}{2}\sinh(2z).
\end{equation*}
This temporally periodic, spatially localized breather is referred to as the Kuznetsov-Ma breather. Its time period reads
\begin{equation*}
T=\frac{2\pi}{\sinh2z}.
\end{equation*}

Under this circumstance, the limit $z \to 0$ transforms the solution into
\begin{equation*}
\psi= \textrm{i}\left(-1+4\frac{1+2\textrm{i}t}{1+4x^2+4t^2}\right)e^{\textrm{i}t}.
\end{equation*}
This solution is a Peregrine soliton, also known as a rogue wave.

Substitution of $z=0$ into Eq.\eqref{eq36} gives rise to an Akhmediev breather that is spatially periodic and temporally localized, its form reads
\begin{equation}
\psi= \textrm{i}\left(\frac{\cos\alpha\cos(2\beta x)-\cos(2\alpha)\cosh( 2\gamma t)-\textrm{i}\sin2\alpha\sinh2\gamma t}{-\cos\alpha\cos(2\beta x)+\cosh(2\gamma t)}\right)e^{\textrm{i}t},
\end{equation}
where
\begin{equation*}
\gamma=-\frac{1}{2}\sin2\alpha, \quad \beta=\sin\alpha.
\end{equation*}
The periods along the spatial coordinate is expressed as
\begin{equation*}
L=\frac{\pi}{\sin\alpha}.
\end{equation*}

While one-solitonic solution of the standard NLSE contains merely one wave, the one-solitonic solution of Eq.\eqref{eq1} is formed by the interaction between two waves. It is symmetric with respect to the time axis. The one-solitonic solution is illustrated in Fig.~1(a). Observing Fig.~1(a), one can see a singularity at the sharp peak, which gives rise to collapse. A sharply rising peak emerges during the collapse of solitons. Figure 1(b) exhibits the spatiotemporal evolution of the Kuznetsov-Ma breather. The behaviour of Akhmediev breather is given in Fig.~1(c). A `quasi-Akhmediev' breather is formed as $z \to 0$. From Fig.~2, one can observe that the wave profile gradually shrinks as $\alpha$ decreases. Unlike the `quasi-Akhmediev' breather of the standard NLSE, Fig.~2(d) exhibits two dominant peaks.

\noindent
{\rotatebox{0}{\includegraphics[width=4.8cm,height=3.8cm]{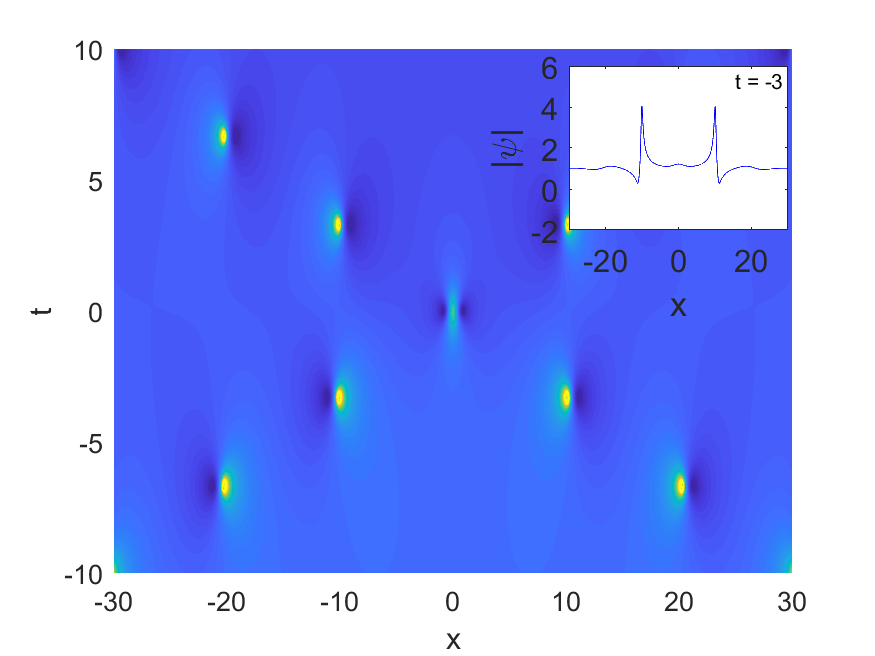}}}
{\rotatebox{0}{\includegraphics[width=4.8cm,height=3.8cm]{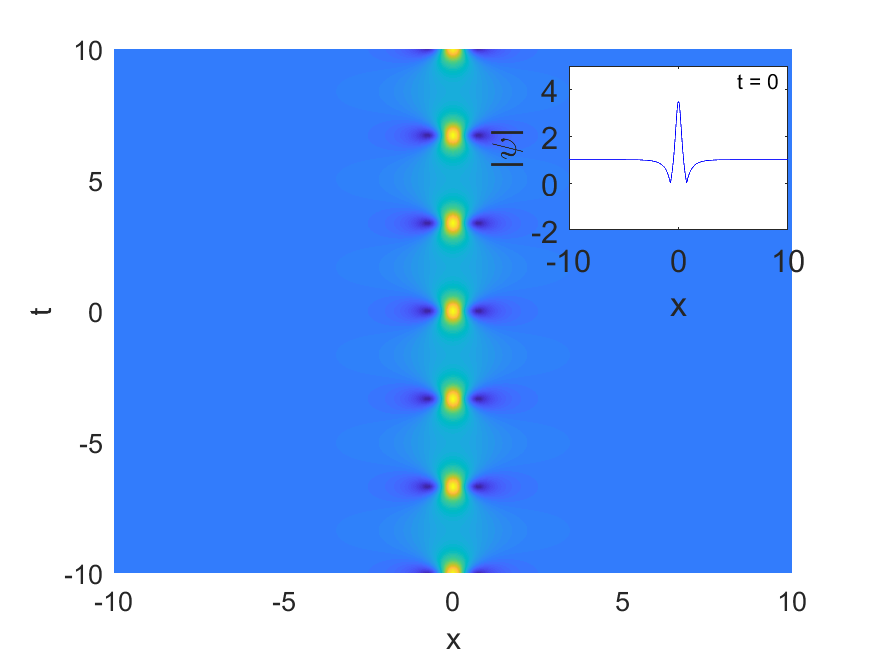}}}
{\rotatebox{0}{\includegraphics[width=4.8cm,height=3.8cm]{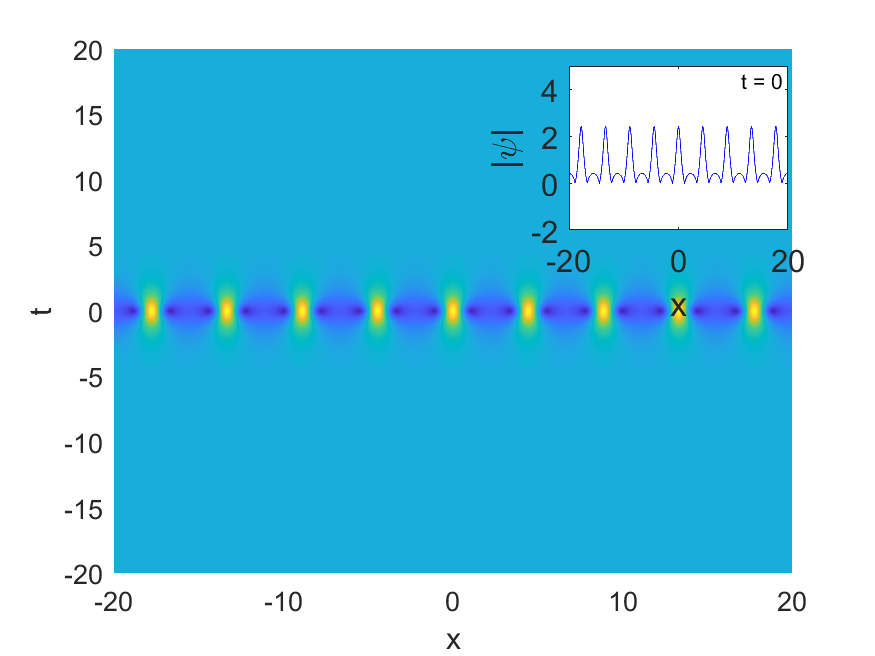}}}\\
$~~~~~~~~~~~~~~~~~~(\textbf{a})
~~~~~~~~~~~~~~~~~~~~~~~~~~~~~~~~~~~(\textbf{b})
~~~~~~~~~~~~~~~~~~~~~~~~~~~~~~~~~~~(\textbf{c})$\\

\noindent { \small \textbf{Figure 1.} One-solitonic solution, Kuznetsov-Ma breather, Akhmediev breather to Eq.\eqref{eq1} in the defocusing regime with parameters: $\textbf{(a)}$: $\alpha=\pi/11$, $z=0.095$, $\mu=0$, $\theta=\pi$, $\textbf{(b)}$: $\alpha=0$, $z=\ln2$, $\mu=0$, $\theta=\pi$, $\textbf{(c)}$: $\alpha=\pi/4$, $z=0$, $\mu=0$, $\theta=\pi$. \\}

\noindent
{\rotatebox{0}{\includegraphics[width=4.8cm,height=3.8cm]{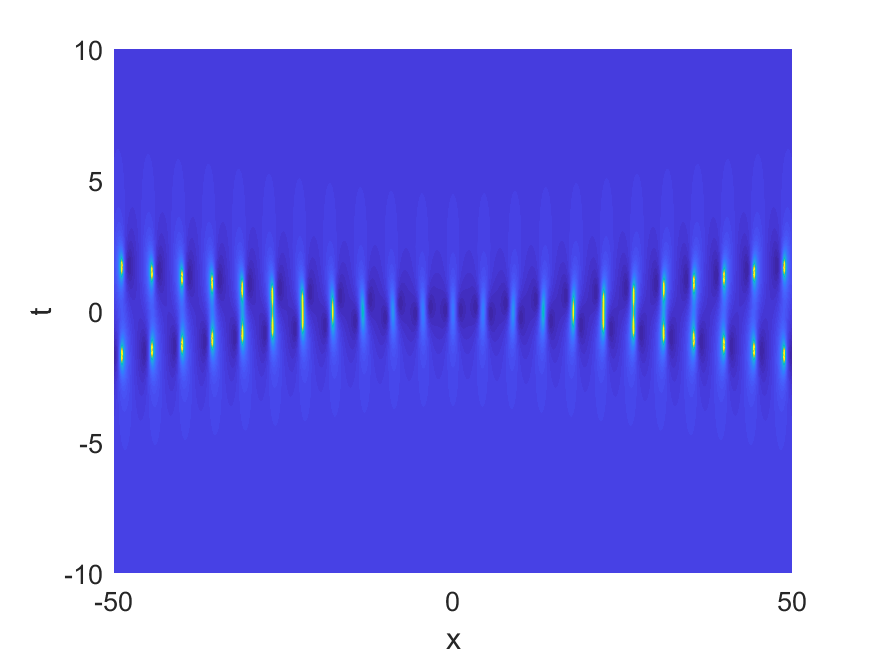}}}
{\rotatebox{0}{\includegraphics[width=4.8cm,height=3.8cm]{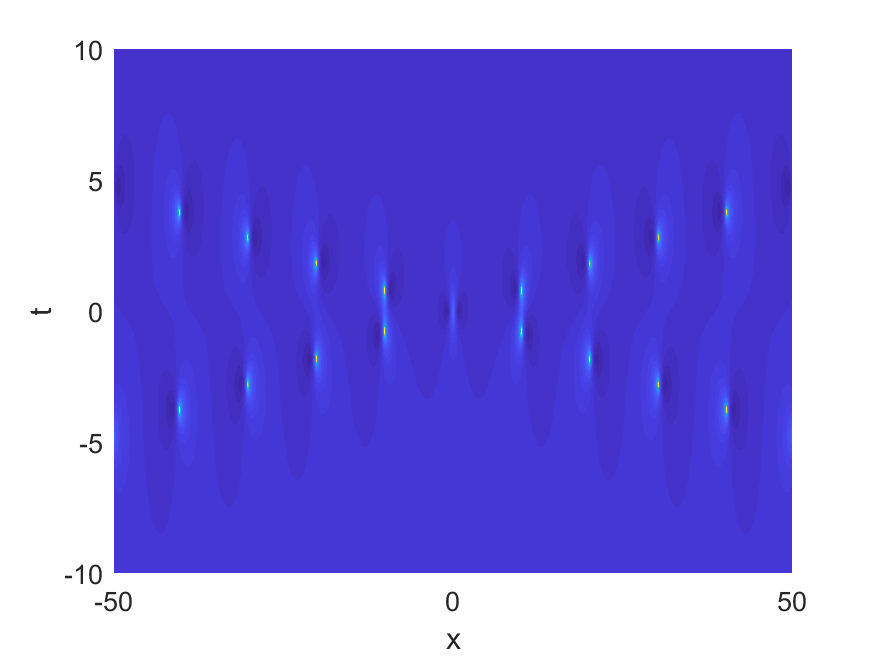}}}
{\rotatebox{0}{\includegraphics[width=4.8cm,height=3.8cm]{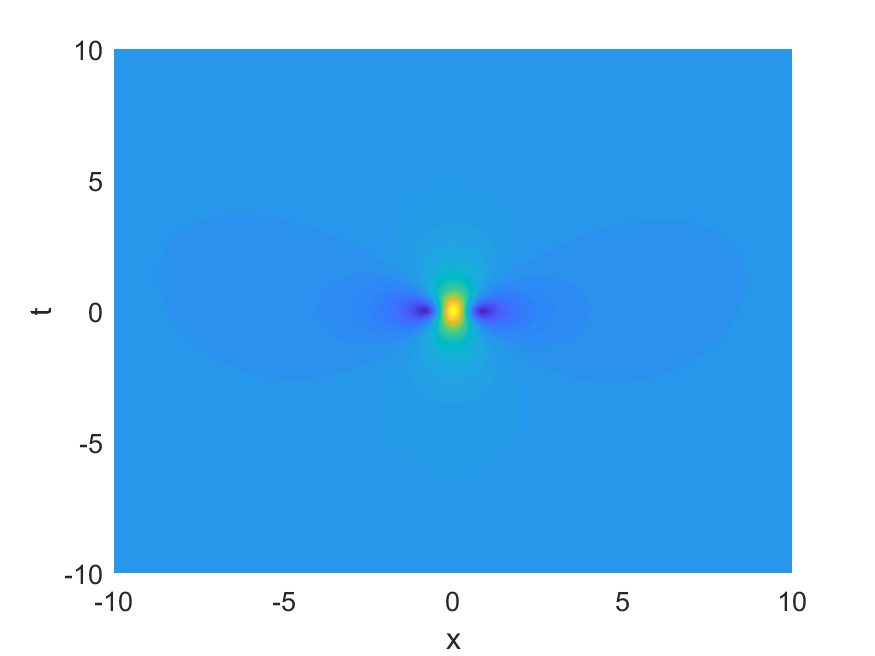}}}\\
$~~~~~~~~~~~~~~~~~~(\textbf{a})
~~~~~~~~~~~~~~~~~~~~~~~~~~~~~~~~~~~(\textbf{b})
~~~~~~~~~~~~~~~~~~~~~~~~~~~~~~~~~~~(\textbf{c})$\\
{\rotatebox{0}{\includegraphics[width=4.8cm,height=3.8cm]{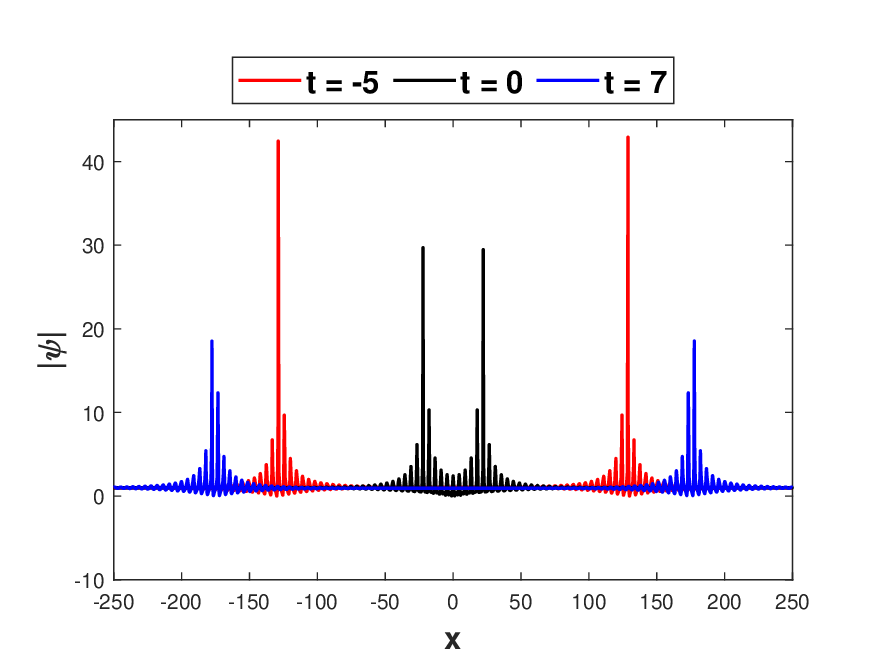}}}
{\rotatebox{0}{\includegraphics[width=4.8cm,height=3.8cm]{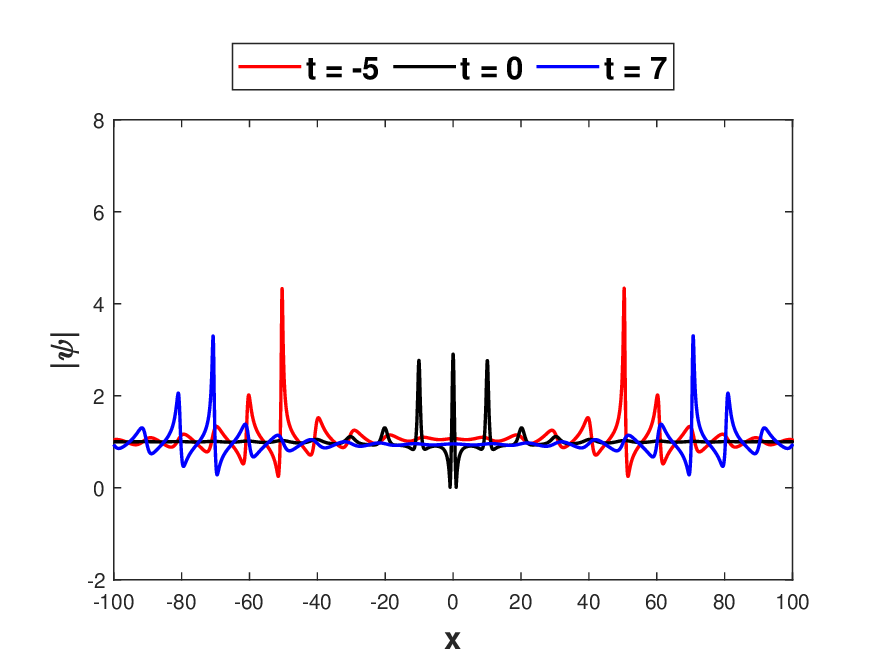}}}
{\rotatebox{0}{\includegraphics[width=4.8cm,height=3.8cm]{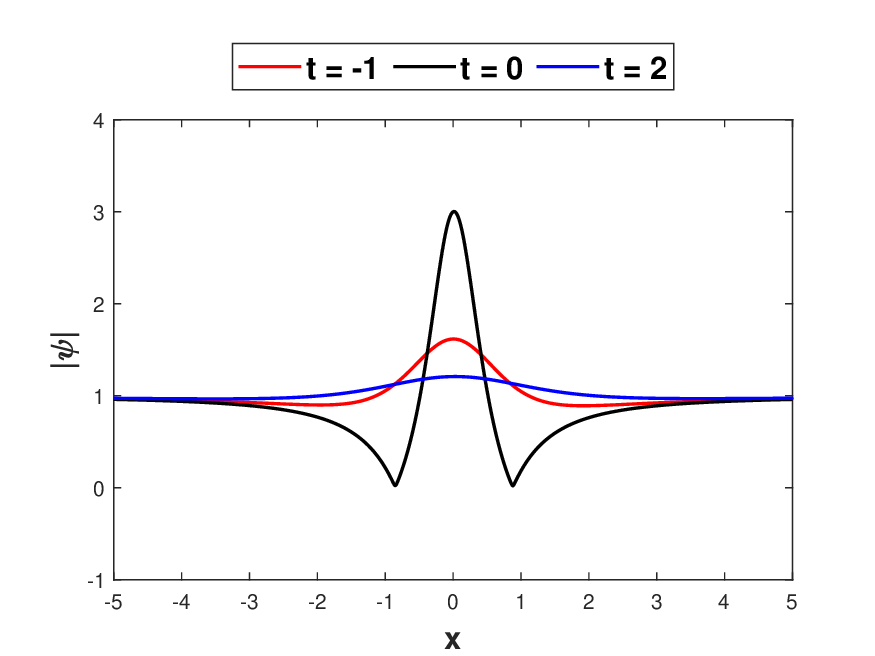}}}\\
$~~~~~~~~~~~~~~~~~~(\textbf{d})
~~~~~~~~~~~~~~~~~~~~~~~~~~~~~~~~~~~(\textbf{e})
~~~~~~~~~~~~~~~~~~~~~~~~~~~~~~~~~~~(\textbf{f})$\\

\noindent { \small \textbf{Figure 2.} Profiles of `quasi-Akhmediev' breather to Eq.\eqref{eq1} in the defocusing regime with parameters:
$\sigma=-1$, $z=0.03$, $\mu=0$, $\theta=\pi$. $\textbf{(a,~d)}$: $\alpha=\pi/4$. $\textbf{(b,~e)}$: $\alpha=\pi/10$. $\textbf{(c,~f)}$: $\alpha=\pi/100$.\\}

With the analytical expression of the one-solitonic solution fully derived, we turn our attention to the construction of two-solitonic solutions below.

\subsection{Two-solitonic solution}
The two-solitonic solution is derived from Eq.\eqref{eq31}. It admits two equivalent expressions. One is formulated in terms of the parameter $\lambda$, which reads
\begin{align}
\psi &= \textrm{i}\left(1 + 2\frac{N_\lambda}{\Delta_\lambda}\right)e^{\textrm{i}t}, \label{eq37} \\
N_\lambda &= \frac{q_{12}q_{21}(-t)\boldsymbol{q}_2^T \boldsymbol{q}_1(-t) + q_{22}q_{11}(-t)\boldsymbol{q}_1^T \boldsymbol{q}_2(-t)}{\lambda_1 + \lambda_2} \nonumber\\
&- \frac{q_{22}q_{21}(-t)\boldsymbol{q}_1^T \boldsymbol{q}_1(-t)}{2\lambda_1}
- \frac{q_{12}q_{11}(-t)\boldsymbol{q}_2^T \boldsymbol{q}_2(-t)}{2\lambda_2}, \nonumber \\
\Delta_\lambda &= \frac{\boldsymbol{q}_1^T(-t)\boldsymbol{q}_1 \boldsymbol{q}_2^T(-t)\boldsymbol{q}_2}{4\lambda_1\lambda_2}
- \frac{\boldsymbol{q}_1^T(-t)\boldsymbol{q}_2 \boldsymbol{q}_2^T(-t)\boldsymbol{q}_1}{(\lambda_1 + \lambda_2)^2}.\nonumber
\end{align}
The other takes the parameterized variable $\xi$ as the argument, given by
\begin{equation}\label{eq38}
\psi = \textrm{i}\left(1 + 2\,\frac{N_\xi}{\Delta_\xi}\right)e^{\textrm{i}t}.
\end{equation}
Here
\begin{align*}
N_\xi &= H_1\big[q_{12}q_{21}(-t)\boldsymbol{q}_2^T \boldsymbol{q}_1(-t) + q_{22}q_{11}(-t)\boldsymbol{q}_1^T \boldsymbol{q}_2(-t)\big] \nonumber \\
& - H_2 q_{22}q_{21}(-t)\boldsymbol{q}_1^T \boldsymbol{q}_1(-t) - H_3 q_{12}q_{11}(-t)\boldsymbol{q}_2^T \boldsymbol{q}_2(-t), \nonumber \\
\Delta_\xi &= H_4 \boldsymbol{q}_1^T(-t)\boldsymbol{q}_1 \boldsymbol{q}_2^T(-t)\boldsymbol{q}_2 - H_5 \boldsymbol{q}_1^T(-t)\boldsymbol{q}_2 \boldsymbol{q}_2^T(-t)\boldsymbol{q}_1.
\end{align*}
The vectors $\boldsymbol{q}_1=(q_{11},~q_{12})^T$ and $\boldsymbol{q}_2=(q_{21},~q_{22})^T$ can be obtained by Eq.\eqref{qn}, with the auxiliary quantities $H_1\sim H_5$ defined as follows:
\begin{align*}
H_1 &= 4\bigl(\cos\alpha_1 \cosh z_1 + \textrm{i} \sin\alpha_1 \sinh z_1\bigr)\bigl(\cos\alpha_2 \cosh z_2 + \textrm{i} \sin\alpha_2 \sinh z_2\bigr) \notag \\
&\quad \cdot \bigl(\cos\alpha_1 \cosh z_1 + \textrm{i}\sin\alpha_1 \sinh z_1 + \cos\alpha_2 \cosh z_2 + \textrm{i} \sin\alpha_2 \sinh z_2\bigr), \notag \\
H_2 &= 2\bigl(\cos\alpha_2 \cosh z_2 + \textrm{i}\sin\alpha_2 \sinh z_2\bigr) \notag \\
&\quad \cdot \bigl(\cos\alpha_1 \cosh z_1 + \textrm{i} \sin\alpha_1 \sinh z_1 + \cos\alpha_2 \cosh z_2 + \textrm{i} \sin\alpha_2 \sinh z_2\bigr)^2, \notag \\
H_3 &= 2\bigl(\cos\alpha_1 \cosh z_1 + \textrm{i} \sin\alpha_1 \sinh z_1\bigr) \notag \\
&\quad \cdot \bigl(\cos\alpha_1 \cosh z_1 + \textrm{i} \sin\alpha_1 \sinh z_1 + \cos\alpha_2 \cosh z_2 + \textrm{i} \sin\alpha_2 \sinh z_2\bigr)^2, \notag \\
H_4 &= \bigl(\cos\alpha_1 \cosh z_1 + \textrm{i} \sin\alpha_1 \sinh z_1 + \cos\alpha_2 \cosh z_2 + \textrm{i} \sin\alpha_2 \sinh z_2\bigr)^2, \notag \\
H_5 &= 4\bigl(\cos\alpha_1 \cosh z_1 + \textrm{i} \sin\alpha_1 \sinh z_1\bigr)\bigl(\cos\alpha_2 \cosh z_2 + \textrm{i} \sin\alpha_2 \sinh z_2\bigr).
\end{align*}

When the parameters $\alpha_1$ and $\alpha_2$ are both nonzero and the group velocities are unequal, the two solitons will inevitably collide. By appropriately choosing the parameters, we can always set the collision time and position to the origin, i.e., $x=0,\,t=0$. Therefore, we directly set $\mu_1=\mu_2=0$ in the subsequent analysis. The sign of $\alpha$ determines the moving direction of the soliton. The two-solitonic solution for solitons moving and colliding in the same direction is shown in Fig.~3(a). The analytical solution intrinsically contains four characteristic wave components. The figure shows the characteristic dynamical behavior of MI for the nonlocal model. Due to the nonlocal coupling of Eq.\eqref{eq1} under time reversal, the fields of the solution for $t<0$ and $t>0$ must satisfy a self-consistency condition simultaneously. This constraint leads to background fields with different levels for $t<0$ and $t>0$.
As the evolution approaches $t=0$, perturbations are amplified, and a series of spatially periodic local intensity maxima emerge in the spatiotemporal transition region, forming a `quasi-Akhmediev' breather sequence. These periodic localized structures are bounded by a V-shaped spatiotemporal envelope and mainly concentrate in the overlapping transition zone between the high background ($t<0$) and the low background after attenuation ($t>0$).

From section \ref{sec2}, we know that the solution is regular if the sum of all $\alpha$ equals $0$. For the two-solitonic case, when we set $\alpha_1 = -\alpha_2 = \alpha$, the resulting solution belongs to the regular two-solitonic branch. This solution comprises two solitons propagating in opposite directions that undergo mutual collision. The regular two-solitonic solution is shown in Fig.~3(b). It is symmetric about the t-axis but not symmetric about the x-axis. The group velocities are
\begin{equation*}
V_{gr1}=-\frac{\sin\alpha\cosh2z_1}{\sinh z_1}, \quad V_{gr2}=\frac{\sin\alpha\cosh2z_2}{\sinh z_2}.
\end{equation*}

Under the symmetric condition $z_1=z_2=z$, the variable expressions contained in $\phi_1$ and $\phi_2$ take the form:
\begin{align*}
&\text{\ae}_1=\text{\ae}_2=\text{\ae}=\cos\alpha\sinh z,\nonumber \\
&\gamma_1=-\gamma_2=\gamma=-\frac{1}{2}\sin(2\alpha)\cosh(2z),\nonumber \\
&\beta_1=-\beta_2=\beta=\sin\alpha\cosh z, \nonumber \\
&\omega_1=\omega_2=\omega=\frac{1}{2}\cos(2\alpha)\sinh(2z).
\end{align*}
In this case (see Fig.~3(c)), the expression for the two-solitonic solution reads:
\begin{align*}
\psi &=\textrm{i}\left(1+4\cos\alpha\cosh z \frac{N_\xi}{\Delta_\xi}\right)e^{\textrm{i}t},\nonumber \\
N_\xi &=(\cosh2z+\cos2\alpha)\big[q_{12}q_{21}(-t)\boldsymbol{q}_2^T \boldsymbol{q}_1(-t) + q_{22}q_{11}(-t)\boldsymbol{q}_1^T \boldsymbol{q}_2(-t)\big]\nonumber \\
&-2\cos\alpha\cosh z (\cos\alpha\cosh z-\textrm{i}\sin\alpha\sinh z)q_{22}q_{21}(-t)\boldsymbol{q}_1^T \boldsymbol{q}_1(-t)\nonumber \\
&-2\cos\alpha\cosh z (\cos\alpha\cosh z+\textrm{i}\sin\alpha\sinh z)q_{12}q_{11}(-t)\boldsymbol{q}_2^T \boldsymbol{q}_2(-t),\nonumber \\
\Delta_\xi &=2\cos^2 \alpha\cosh^2 z\boldsymbol{q}_1^T(-t)\boldsymbol{q}_1 \boldsymbol{q}_2^T(-t)\boldsymbol{q}_2
-(\cos2\alpha+\cosh2z)\boldsymbol{q}_1^T(-t)\boldsymbol{q}_2 \boldsymbol{q}_2^T(-t)\boldsymbol{q}_1. \label{eq42}
\end{align*}

\noindent
{\rotatebox{0}{\includegraphics[width=4.8cm,height=3.8cm]{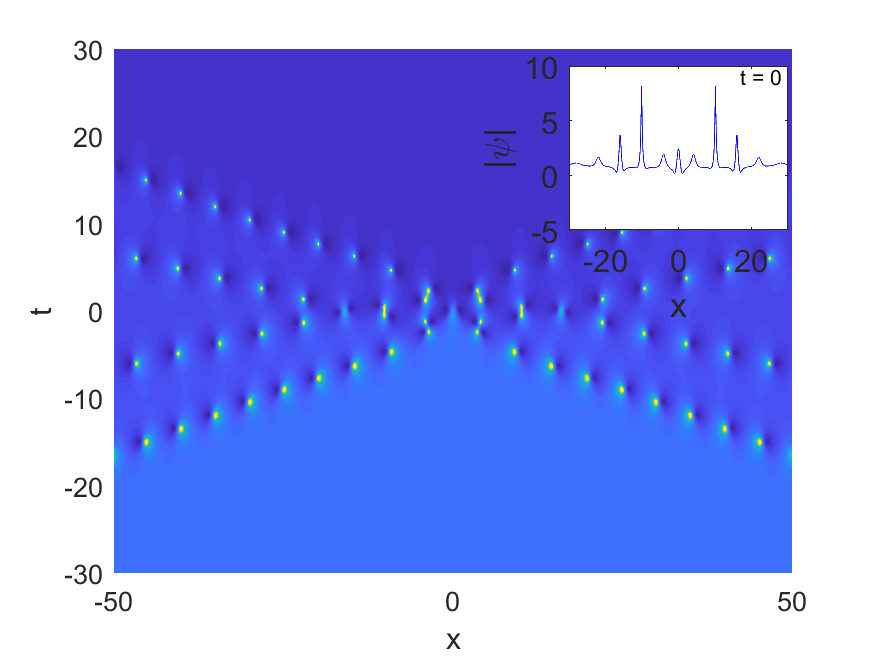}}}
{\rotatebox{0}{\includegraphics[width=4.8cm,height=3.8cm]{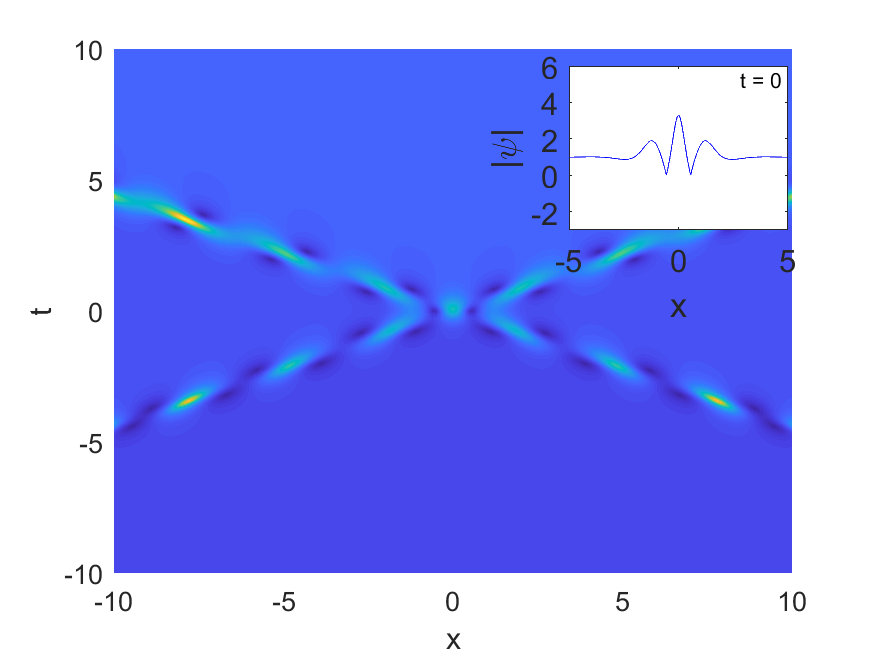}}}
{\rotatebox{0}{\includegraphics[width=4.8cm,height=3.8cm]{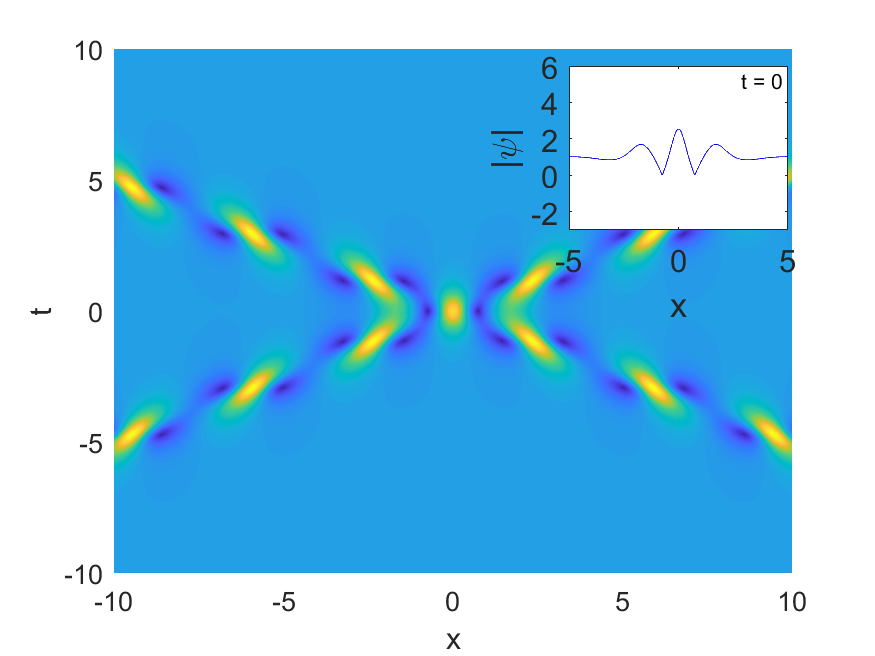}}}\\
$~~~~~~~~~~~~~~~~~~(\textbf{a})
~~~~~~~~~~~~~~~~~~~~~~~~~~~~~~~~~~~(\textbf{b})
~~~~~~~~~~~~~~~~~~~~~~~~~~~~~~~~~~~(\textbf{c})$\\

\noindent { \small \textbf{Figure 3.} Two-solitonic solution in the defocusing regime. $\textbf{(a)}$: General two-solitonic solution with singularity $\alpha_1=\pi/6,~\alpha_2=\pi/5$, $R_1=1.1,~ R_2=1.2$, $\mu_1=\mu_2=0$, $\theta_1=\theta_2=0$. $\textbf{(b)}$: Regular two-solitonic solution in the asymmetric case $\alpha_1=\pi/4,~\alpha_2=-\pi/4$, $z_1=0.916,~z_2=1.03$, $\mu_1=\mu_2=0$, $\theta_1=\theta_2=0$. $\textbf{(c)}$: Regular two-solitonic solution in the symmetric case $\alpha_1=\pi/4,~\alpha_2=-\pi/4$, $z_1=z_2=\ln2$, $\mu_1=\mu_2=0$, $\theta_1=\theta_2=0$.\\}

It becomes a double Akhmediev breather when $z_1=z_2=0$. As shown in Fig.~4(a), the two breathers collide at $x=0,~t=0$ and form a single peak, and the overall shape exhibits an X-type structure. In the central collision region, strong nonlinear coupling distorts the field profile. After interaction, the waveforms of both Akhmediev breathers are fully recovered. The solution represents double Kuznetsov-Ma breather when all poles lie on the real axis, i.e., $\alpha_1=\alpha_2=0$. The breathers oscillate periodically along the $t$ axis and remain spatially localized near $x=0$, generating two vertically aligned elliptical closed-loop structures, as shown in Fig.~4(b). Figure 4(c) shows a two-solitonic state formed by a Kuznetsov-Ma breather and an Akhmediev breather. It exhibits a cross-shaped pattern. The interaction between an Akhmediev breather and a soliton is presented in Fig.~5. Before collision, paired complex poles generate alternating bright and dark fringes of the intensity $|\psi(x,t)|$ on the real $(x,t)$ plane. When the collision occurs near $t\approx0$, the modulation of bright-dark oscillations in the core interaction region is suppressed. After collision, the periodic fringes characteristic of the Akhmediev breather are fully restored.

\noindent
{\rotatebox{0}{\includegraphics[width=4.8cm,height=3.8cm]{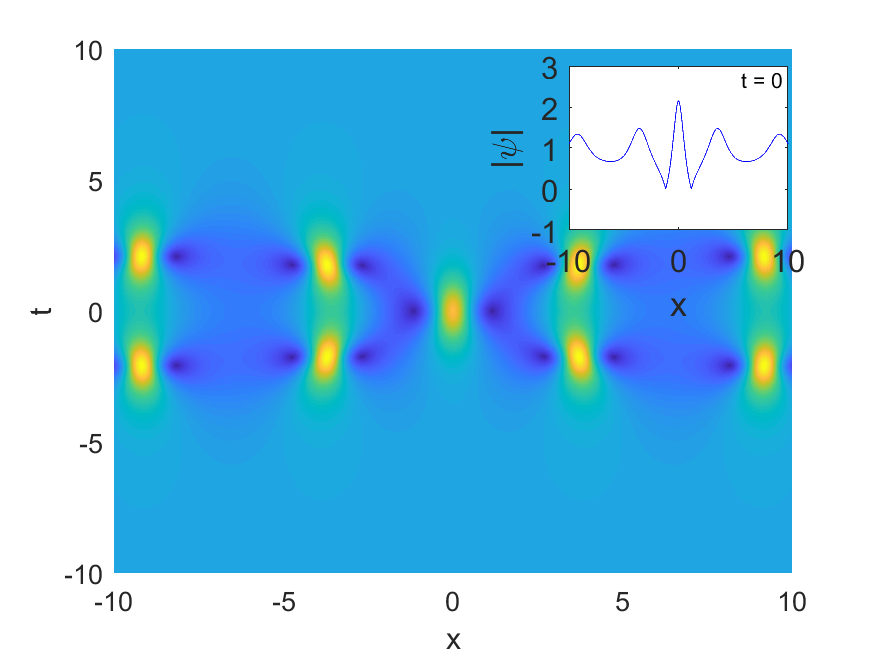}}}
{\rotatebox{0}{\includegraphics[width=4.8cm,height=3.8cm]{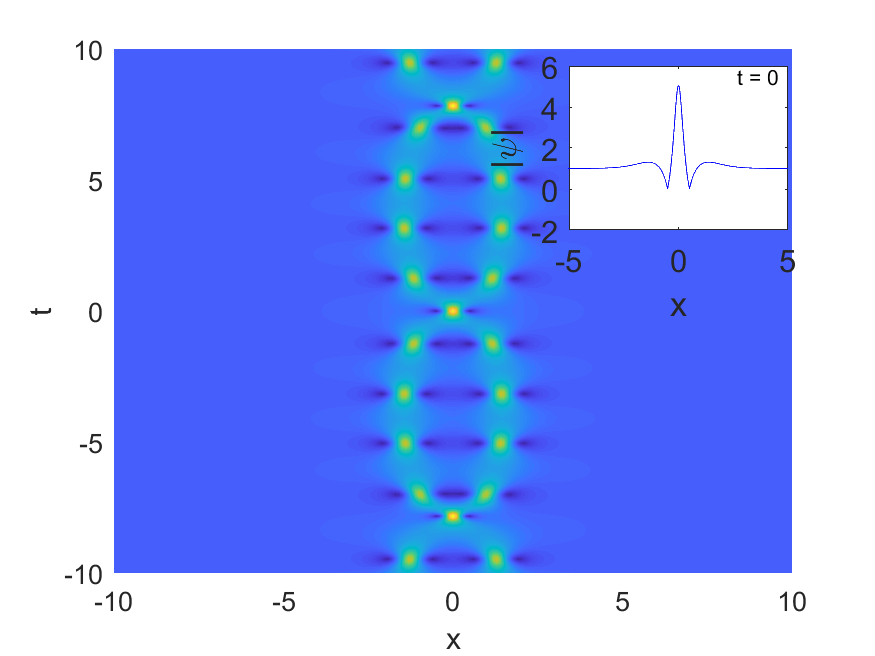}}}
{\rotatebox{0}{\includegraphics[width=4.8cm,height=3.8cm]{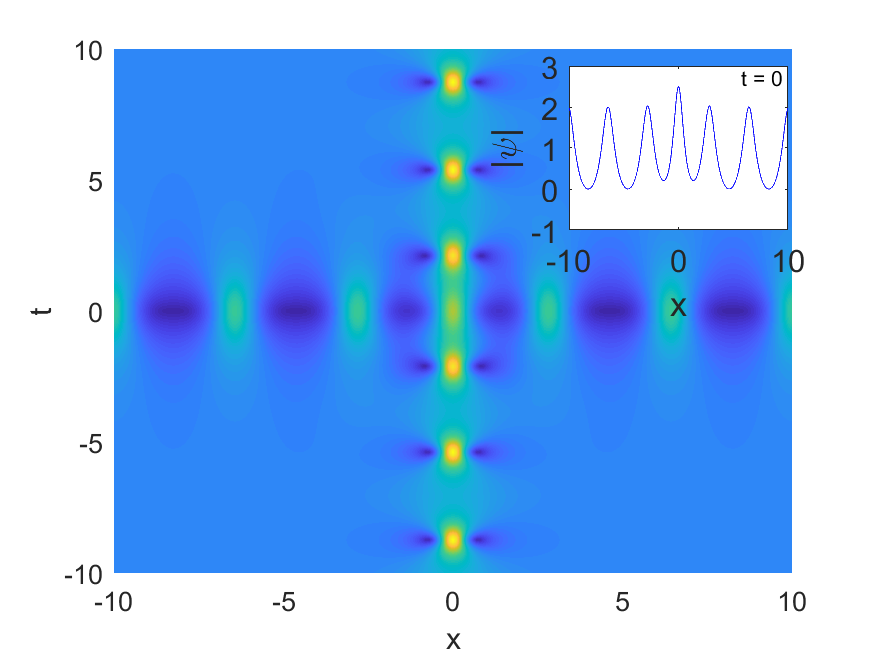}}}\\
$~~~~~~~~~~~~~~~~~~(\textbf{a})
~~~~~~~~~~~~~~~~~~~~~~~~~~~~~~~~~~~(\textbf{b})
~~~~~~~~~~~~~~~~~~~~~~~~~~~~~~~~~~~(\textbf{c})$\\

\noindent { \small \textbf{Figure 4.} Evolution of breather structures in the defocusing regime. $\textbf{(a)}$: Double Akhmediev breather with $\alpha_1=\pi/6,~\alpha_2=\pi/4$, $z_1=z_2=0$, $\mu_1=\mu_2=0$, $\theta_1=\theta_2=0$. $\textbf{(b)}$: Double Kuznetsov-Ma breather with $\alpha_1= \alpha_2=0$, $z_1=0.916,~z_2=1.03$, $\mu_1=\mu_2=0$, $\theta_1=\theta_2=0$. $\textbf{(c)}$: Superposition of the Kuznetsov-Ma and Akhmediev breather with $\alpha_1=0,~\alpha_2=\pi/3$, $z_1=\ln2,~z_2=0$, $\mu_1=\mu_2=0$, $\theta_1=\theta_2=0$.\\}

\noindent
{\rotatebox{0}{\includegraphics[width=4.8cm,height=3.8cm]{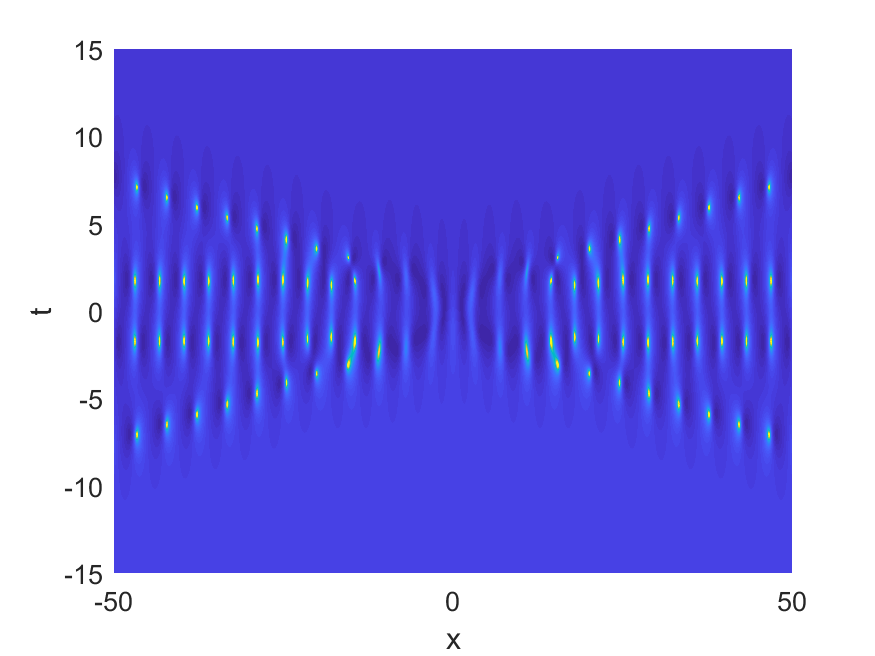}}}
{\rotatebox{0}{\includegraphics[width=4.8cm,height=3.8cm]{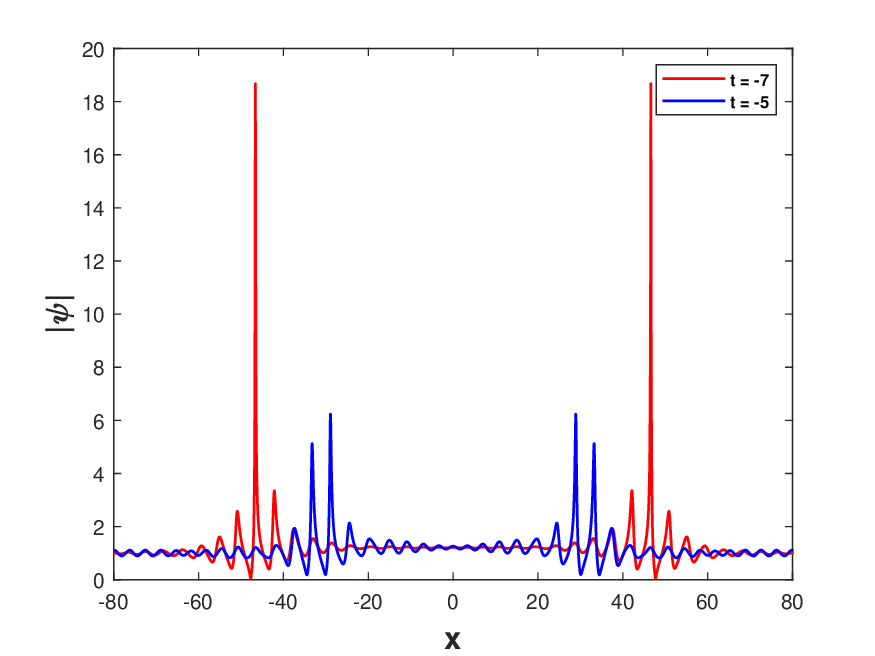}}}
{\rotatebox{0}{\includegraphics[width=4.8cm,height=3.8cm]{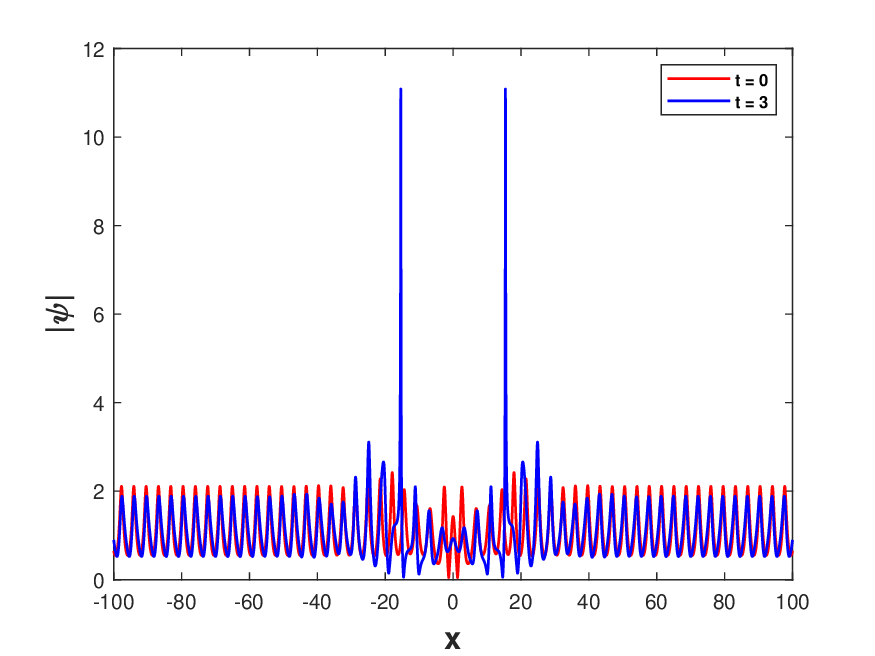}}}\\
$~~~~~~~~~~~~~~~~~~(\textbf{a})
~~~~~~~~~~~~~~~~~~~~~~~~~~~~~~~~~~~(\textbf{b})
~~~~~~~~~~~~~~~~~~~~~~~~~~~~~~~~~~~(\textbf{c})$\\

\noindent { \small \textbf{Figure 5.} Combination of Akhmediev breather and a soliton in the defocusing regime with parameters: $\alpha_1=\pi/4,~ \alpha_2=\pi/3$, $z_1=0.095,~z_2=0$, $\mu_1=\mu_2=0$, $\theta_1=\theta_2=0$. $\textbf{(a)}$: density plot. $\textbf{(b)}$: profiles of $|\psi|$ at $t=-7$ and $t=-5$. $\textbf{(c)}$: profiles of $|\psi|$ at $t=0$ and $t=3$.\\}

\section{Soliton Annihilation}\label{sec4}

We now analyze a special limiting parameter case defined by
\begin{equation}
z_1 = z_2 = 0,\quad \alpha_1 = -\alpha_2 = \alpha.
\end{equation}

In the following, all computations are performed under the \(\lambda\)-parametrization. Under the above parameter constraints, the spectral eigenvalues satisfy
\begin{equation*}
\lambda_1 = \lambda_2 = \cos\alpha.
\end{equation*}
Making use of Eq.\eqref{eq37}, we compute the numerator term
{\small
\[
N_\lambda = \frac{q_{12}q_{21}(-t)\boldsymbol{q}_2^T\!\boldsymbol{q}_1(-t) + q_{22}q_{11}(-t)\boldsymbol{q}_1^T\!\boldsymbol{q}_2(-t) - q_{22}q_{21}(-t)\boldsymbol{q}_1^T\!\boldsymbol{q}_1(-t)-q_{12}q_{11}(-t)\boldsymbol{q}_2^T\!\boldsymbol{q}_2(-t)}{2\cos\alpha} \equiv 0,
\]}
along with the denominator expression
\begin{equation}
\Delta_\lambda = \frac{[q_{11}q_{22} - q_{12}q_{21}][q_{11}(-t)q_{22}(-t)-q_{12}(-t)q_{21}(-t)]}{4\cos^2 \alpha}.
\end{equation}
We further obtain the associated wave parameters:
\begin{align*}
\text{\ae}_1 &= \text{\ae}_2 = 0,~~ \omega_1= -\omega_2 = \omega=\frac{1}{2}\sin 2\alpha, \nonumber\\
\gamma_1 &= \gamma_2 = 0,~~ \beta_1= -\beta_2= \beta  = \sin\alpha.
\end{align*}
The phase arguments read
\begin{gather*}
u_1 = \mu_1/2,\quad v_1 = \beta x+\omega t+\theta_1/2, \nonumber\\
u_2 = \mu_2/2,\quad v_2 = -\beta x - \omega t+\theta_2/2.
\end{gather*}
For compact notation, we introduce the combined phase quantities
\begin{equation}
\mu_1 \pm \mu_2 = \mu^\pm,\qquad \theta_1 \pm \theta_2 = \theta^\pm.
\end{equation}
By calculation, we get
\begin{equation}
q_{11}q_{22} - q_{12}q_{21} = -4\textrm{i}\sin\alpha\left(\cos\frac{\theta^+}{2}\sinh\frac{\mu^+}{2}+ \textrm{i}\sin\frac{\theta^+}{2}\cosh\frac{\mu^+}{2}\right).
\end{equation}
Substituting this result into $\Delta_\lambda$, we arrive at the final closed-form expression
\begin{equation}
\Delta_\lambda=-\frac{4\sin^2\alpha}{\cos^2\alpha}\left(\cos\frac{\theta^+}{2}\sinh\frac{\mu^+}{2}
+\textrm{i}\sin\frac{\theta^+}{2}\cosh\frac{\mu^+}{2}\right)^2.
\end{equation}
In this limiting configuration, the denominator $\Delta_\lambda$ is a pure constant independent of both $x$ and $t$. If we discard the special case \(\alpha=0\), then \(\Delta \neq 0\) whenever \(\theta^+ \neq 0\) or \(\mu^+ \neq 0\). The identity \(N_\lambda \equiv 0\) implies that for generic parameters satisfying \(\theta^+ \neq 0\), two Akhmediev breathers carrying opposite \(\alpha\) values undergo full mutual annihilation.

\section{Superregular two-solitonic solution}\label{sec5}

Two solitons with opposite angular parameter $\alpha$, namely $\alpha_1=-\alpha_2=\alpha$, can fully cancel each other. This key property allows us to build a special important class of $2N$-solitonic solutions. These solutions describe the evolution of tiny perturbations on the condensate. They are a subset of regular solutions. We call them superregular solitonic solutions for the reverse-time nonlocal NLSE. We start with the simple case $N=1$. We assume $z_1$ and $z_2$ are close to $0$:
\begin{equation}\label{eq43}
z_1 \simeq \varepsilon,\qquad z_2 \simeq a\varepsilon.
\end{equation}
We set $\varepsilon>0$, $a>0$, $|\varepsilon|\ll1$. We name this group of poles a reverse-time superregular pair. We use $\delta\psi$ to represent the variational deviation of the full solution from the condensate background at annihilation time:
\begin{equation}
\psi = \textrm{i}e^{\textrm{i}t} + \delta\psi.
\end{equation}
All subsequent calculations are performed under the $\lambda$ variable environment. The expression of \(\delta\psi\) is given by:
\begin{equation*}
\delta\psi=2\textrm{i}\frac{\delta N_\lambda}{\widetilde{\Delta_\lambda}}e^{\textrm{i}t}.
\end{equation*}
Here, \(\delta N_\lambda\) denotes the variation of \(N_\lambda\), and \(\widetilde{\Delta}_\lambda\) stands for the modified denominator. In what follows, we explicitly solve for these two quantities under the defocusing regime.

For the defocusing regime, it follows that
\begin{align*}
\lambda_1 & \approx\cos\alpha+\textrm{i}\varepsilon\sin\alpha, \\
\lambda_2 & \approx\cos\alpha-\textrm{i}a\varepsilon\sin\alpha.
\end{align*}
Then
\begin{equation*}
\lambda_1+\lambda_2=2\cos\alpha+\textrm{i}(1-a)\varepsilon\sin\alpha.
\end{equation*}
Hence
\begin{align}
\widetilde{\Delta}_\lambda &= \frac{\big(\widetilde{q}_{11}\widetilde{q}_{22} - \widetilde{q}_{12}\widetilde{q}_{21}\big)\big(\widetilde{q}_{11}(-t)\widetilde{q}_{22}(-t) - \widetilde{q}_{12}(-t)\widetilde{q}_{21}(-t)\big)}{4\cos^2 \alpha + 2\textrm{i}(1-a)\varepsilon\sin2\alpha}, \\
\delta N_\lambda &= \frac{\textrm{i}(1+a)\varepsilon\sin\alpha}{4\cos^2 \alpha}\Big(
\widetilde{q}_{22}\widetilde{q}_{21}(-t)\widetilde{\boldsymbol{q}}_{1}^T\widetilde{\boldsymbol{q}}_{1}(-t)
- \widetilde{q}_{12}\widetilde{q}_{11}(-t)\widetilde{\boldsymbol{q}}_{2}^T\widetilde{\boldsymbol{q}}_{2}(-t)\Big).
\end{align}
At this point,
\begin{align*}
\beta_1 &=-\beta_2=\beta=\sin\alpha,\nonumber \\
\text{\ae}_1 &=\varepsilon\cos\alpha=\varepsilon\varsigma,\nonumber \\
\text{\ae}_2 &=a\varepsilon\cos\alpha=a\varepsilon\varsigma.
\end{align*}
Here $\varsigma=\cos\alpha$. Let \(t=0\) and \(\mu_1=\mu_2=0\); then we have
\begin{equation*}
\phi_1=\textrm{i}\beta x+\varepsilon\varsigma x+\textrm{i}\frac{\theta_1}{2},\quad \phi_2=-\textrm{i}\beta x+a\varepsilon\varsigma x+\textrm{i}\frac{\theta_2}{2}.
\end{equation*}
Assuming \(\theta^+ \neq 0\), we compute \(\widetilde{\Delta}_\lambda\) and $\delta N_\lambda$ as follows:
\begin{align*}
\widetilde{\Delta}_\lambda &= -4 \cdot\frac{\sin^2 \alpha \big(\cos\frac{\theta^+}{2}\sinh\big(\varepsilon(1+a)\varsigma x\big)
+ \textrm{i}\sin\frac{\theta^+}{2}\cosh\big(\varepsilon(1+a)\varsigma x\big)\big)^2}{\cos^2 \alpha + \textrm{i}(1-a)\varepsilon\sin\alpha\cos\alpha},\\
\delta N_\lambda &=\frac{2\textrm{i}(1+a)\varepsilon\sin\alpha}{\cos^2\alpha}\sin^2\alpha  \Big[
\cos(2\beta x - \theta_2)\cosh(2\alpha\varepsilon\varsigma x)- \cos(2\beta x + \theta_1)\cosh(2\varepsilon\varsigma x)\nonumber \\
&- \textrm{i}\sin(2\beta x - \theta_2)\sinh(2a\varepsilon\varsigma x)- \textrm{i}\sin(2\beta x + \theta_1)\sinh(2\varepsilon\varsigma x)\Big].
\end{align*}
In the symmetric regime \(a=1\), setting \(\theta^+=\pi\), the expression for $\delta\psi$ reads:
\begin{equation}
\delta\psi\approx -4\varepsilon\frac{\sin\alpha\cos(2\beta x-\theta_2)}{\cosh(2\varepsilon\varsigma x)}e^{\textrm{i}t}.
\end{equation}
For \(t \neq 0\), the above formula becomes:
\begin{equation}
\delta\psi\approx -4\varepsilon\frac{(1-\cos\alpha\cos(\alpha+2\omega t))\cos(2\beta x-\theta_2)}{\sin\alpha\cosh(2\varepsilon\varsigma x)}e^{\textrm{i}t}.
\end{equation}

The derived superregular two-solitonic solution possesses the following features. At $t=0$, a small localized initial perturbation generates a pair of solitons propagating in opposite directions with large group velocities. The two solitons are symmetric for $a=1$. Figure 6(a) and (b) display the typical propagation dynamics of superregular two-solitonic solutions under different parameters. A solution close to the superregular form is shown in Fig. 6(c). For $z_1=z_2$, degenerate characteristic roots make the four waves pairwise coalesce, producing straight X-shaped propagation in the $x$--$t$ plane. All components collide elastically at the origin, with interference localized near the collision point. When $z_1\neq z_2$, degeneracy is lifted, and the four waves separate completely. The four-wave interaction yields curved trajectories and widespread interference fringes across the spacetime domain. Small perturbations on the initial background are shown in Fig.~7.

\noindent
{\rotatebox{0}{\includegraphics[width=4.8cm,height=3.8cm]{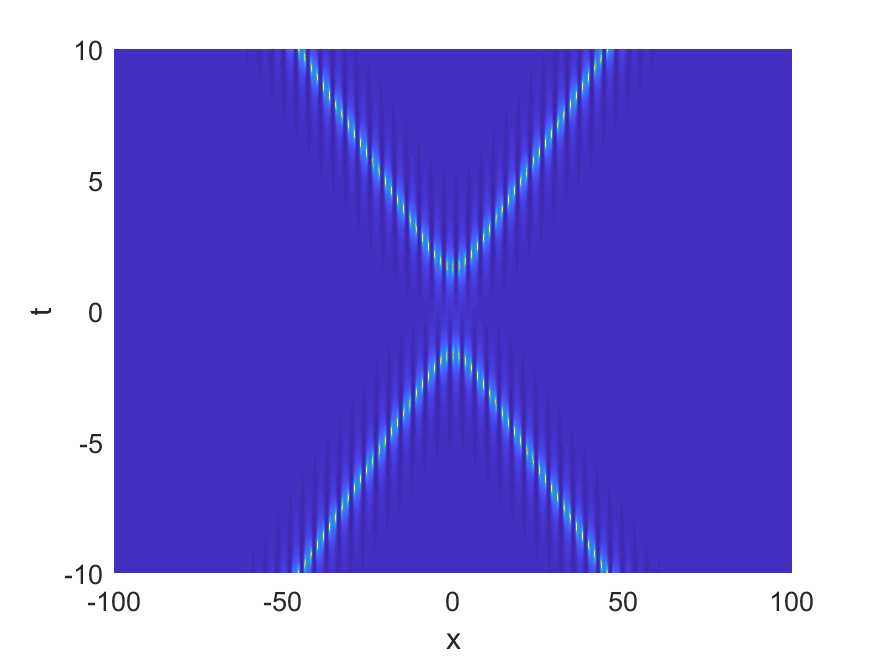}}}
{\rotatebox{0}{\includegraphics[width=4.8cm,height=3.8cm]{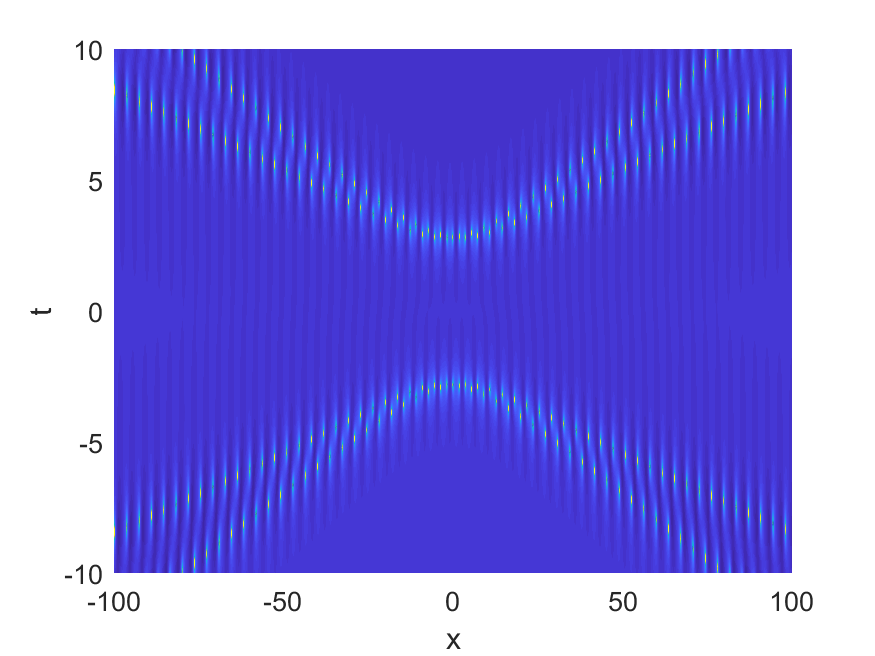}}}
{\rotatebox{0}{\includegraphics[width=4.8cm,height=3.8cm]{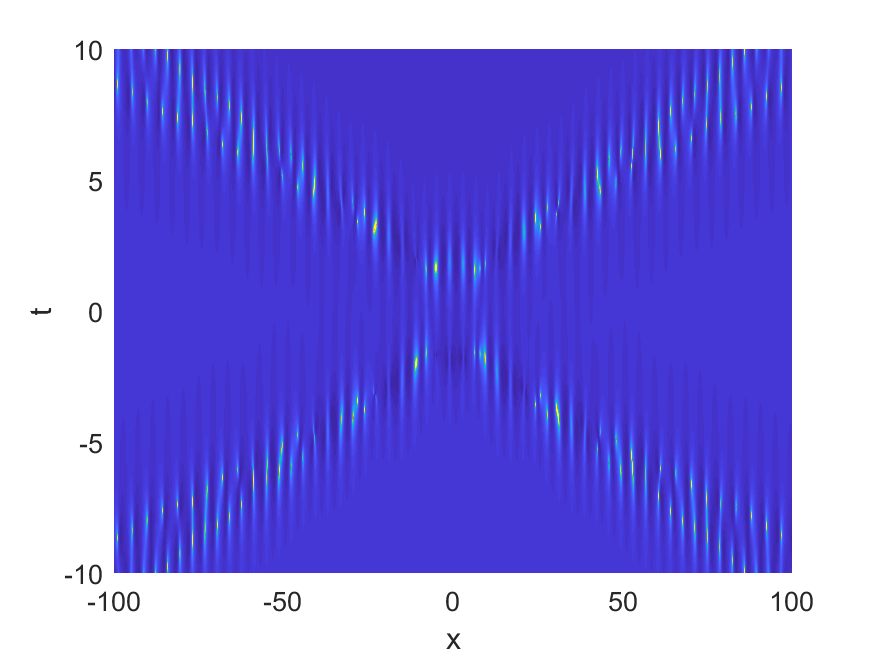}}}\\
$~~~~~~~~~~~~~~~~~~(\textbf{a})
~~~~~~~~~~~~~~~~~~~~~~~~~~~~~~~~~~~(\textbf{b})
~~~~~~~~~~~~~~~~~~~~~~~~~~~~~~~~~~~(\textbf{c})$\\
{\rotatebox{0}{\includegraphics[width=4.8cm,height=3.8cm]{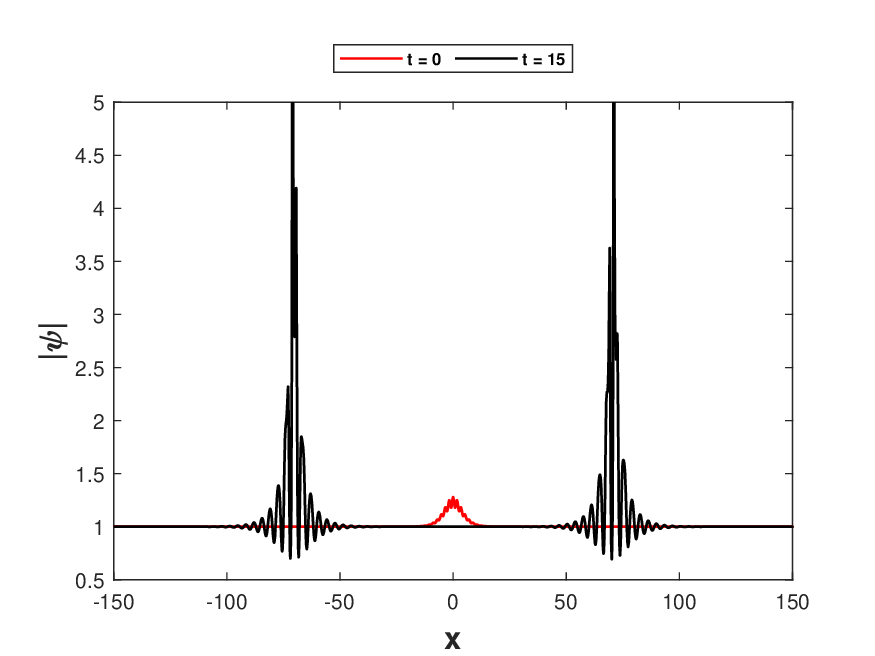}}}
{\rotatebox{0}{\includegraphics[width=4.8cm,height=3.8cm]{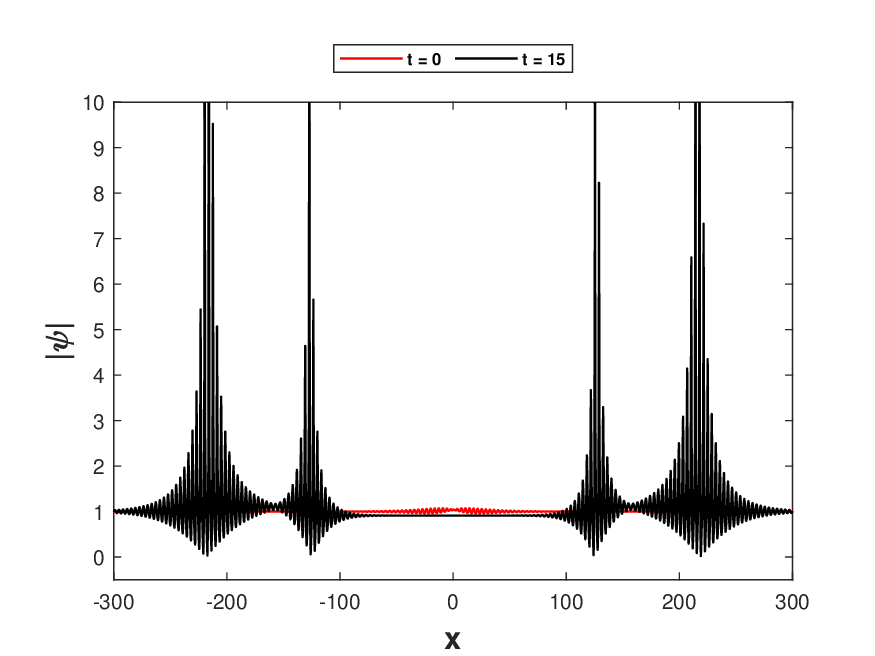}}}
{\rotatebox{0}{\includegraphics[width=4.8cm,height=3.8cm]{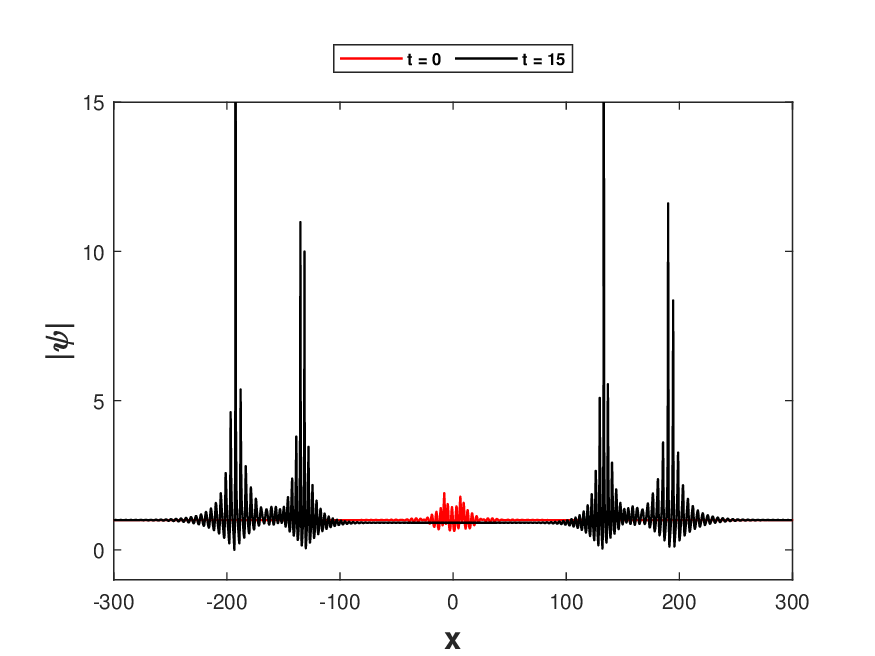}}}\\
$~~~~~~~~~~~~~~~~~~(\textbf{d})
~~~~~~~~~~~~~~~~~~~~~~~~~~~~~~~~~~~(\textbf{e})
~~~~~~~~~~~~~~~~~~~~~~~~~~~~~~~~~~~(\textbf{f})$\\

\noindent { \small \textbf{Figure 6.} Profiles of the superregular two-solitonic solutions and a near-superregular counterpart at moments of $t=0$ and $t=15$ with parameters: $\alpha_1=-\alpha_2=\pi/3$, $z_1=z_2=0.182$, $\theta_1=\theta_2=\pi/2$, $\mu_1=\mu_2=0$ (left), $\alpha_1=-\alpha_2=\pi/3$, $z_1=0.095,~z_2=0.049$, $\theta_1=\theta_2=\pi/2$, $\mu_1=\mu_2=0$ (middle) and $\alpha_1=\pi/3,~\alpha_2=-\pi/4$, $z_1=0.095,~z_2=0.049$, $\theta_1=\theta_2=\pi/2$, $\mu_1=\mu_2=0$ (right).\\}

\noindent
{\rotatebox{0}{\includegraphics[width=4.8cm]{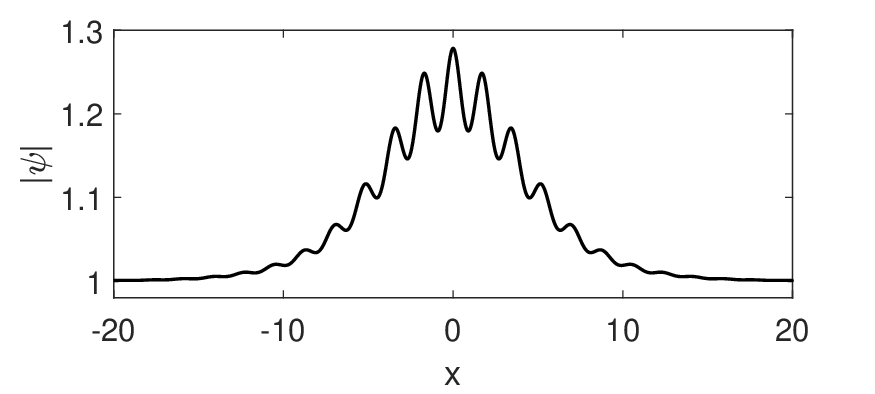}}}
{\rotatebox{0}{\includegraphics[width=4.8cm]{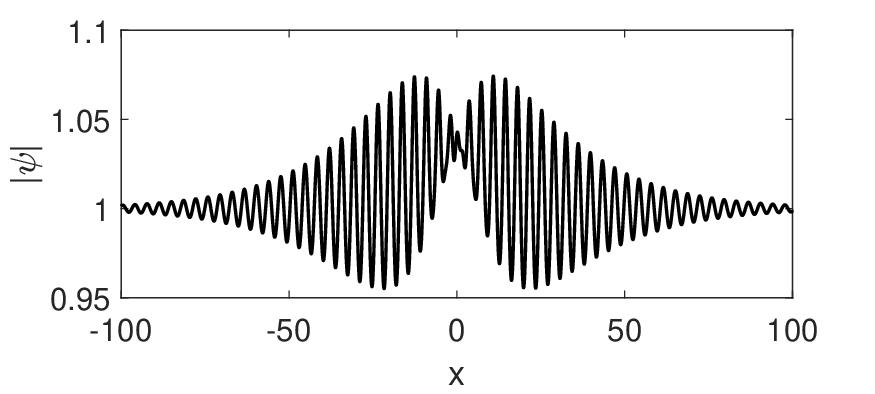}}}
{\rotatebox{0}{\includegraphics[width=4.8cm]{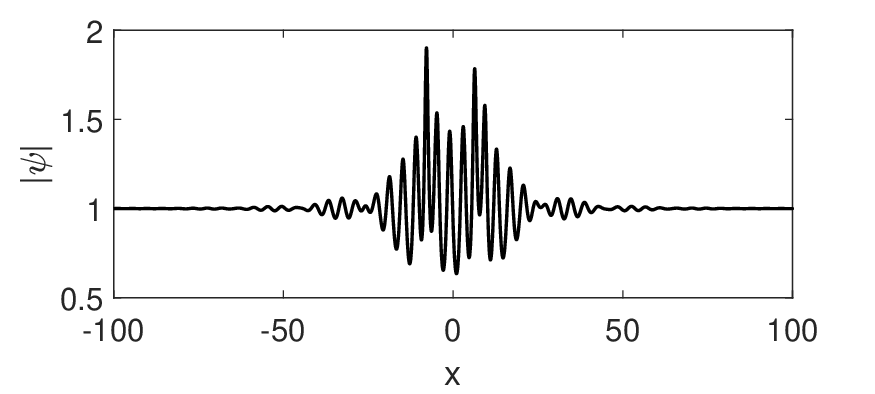}}}\\
$~~~~~~~~~~~~~~~~~~(\textbf{a})
~~~~~~~~~~~~~~~~~~~~~~~~~~~~~~~~~~~(\textbf{b})
~~~~~~~~~~~~~~~~~~~~~~~~~~~~~~~~~~~(\textbf{c})$\\

\noindent { \small \textbf{Figure 7.} Enlarged view of the small perturbation at $t=0$ in Fig.~6.\\}

\begin{rem}
In this work, we have verified that reverse-time nonlocal NLSE \eqref{eq1} in the defocusing regime could support the superregular solitonic solutions, and describe the nonlinear stage of the modulation instability.
For the reverse-time nonlocal NLSE \eqref{eq1} in the focusing regime,
its superregular solitonic solutions can also be obtained by choosing the suitable plane wave solution
in a similar way. Therefore, the detailed derivation and explicit forms of these solutions in the focusing regime are omitted.
\end{rem}

\section{Numerical simulations}\label{sec6}
In this section, numerical simulations are employed to verify the accuracy of the exact solutions, with the one-solitonic solution given by Eq.\eqref{eq36} as an example. To carry out independent numerical verification, we rewrite Eq.\eqref{eq1} into an equivalent coupled system of equations:
\begin{equation*}
\textrm{i}\partial_t \Psi_+ +\frac{1}{2}(\partial_x)^2 \Psi_+ +\sigma(\Psi_+)^2 \Psi_- =0,~~
\textrm{i}\partial_t \Psi_- -\frac{1}{2}(\partial_x)^2 \Psi_- -\sigma(\Psi_-)^2 \Psi_+ =0,
\end{equation*}
where $\Psi_+(x,t)=\psi(x,t)$ and $\Psi_-(x,t)=\psi(x,-t)$. Numerical solutions are obtained via the split-step Fourier method (SSFM), and compared with the analytical solutions. Strang splitting is adopted for numerical propagation, where the equation is decomposed into linear dispersive and nonlinear components. The linear part is solved exactly in Fourier space, while the nonlinear part is integrated using the fourth-order Runge-Kutta (RK4) method. Figures 8(a) and (b) present cross-sectional comparisons between the numerical and analytical solutions. High consistency is found in the peak position, waveform width and overall profile, confirming the validity of Eq.\eqref{eq36}. Nevertheless, detector noise and environmental perturbations cannot be avoided in practical optical measurements. The robustness of the analytical solution in noisy environments is tested by adding additive complex Gaussian white noise to the cross section. The relative noise level is $2\times10^{-2}$, corresponding to $2\%$ of the peak amplitude. Figure 8(b) presents a comparison between the exact solution and the noisy solution for the cross section at $t=1$. The results show that the main waveform structures of the one-solitonic solution (peak position, oscillation period, spatial decay) remain clearly identifiable under the $2\%$ noise level, which demonstrates good robustness of this analytical solution against measurement noise.

\noindent
{\rotatebox{0}{\includegraphics[width=4.8cm,height=3.8cm]{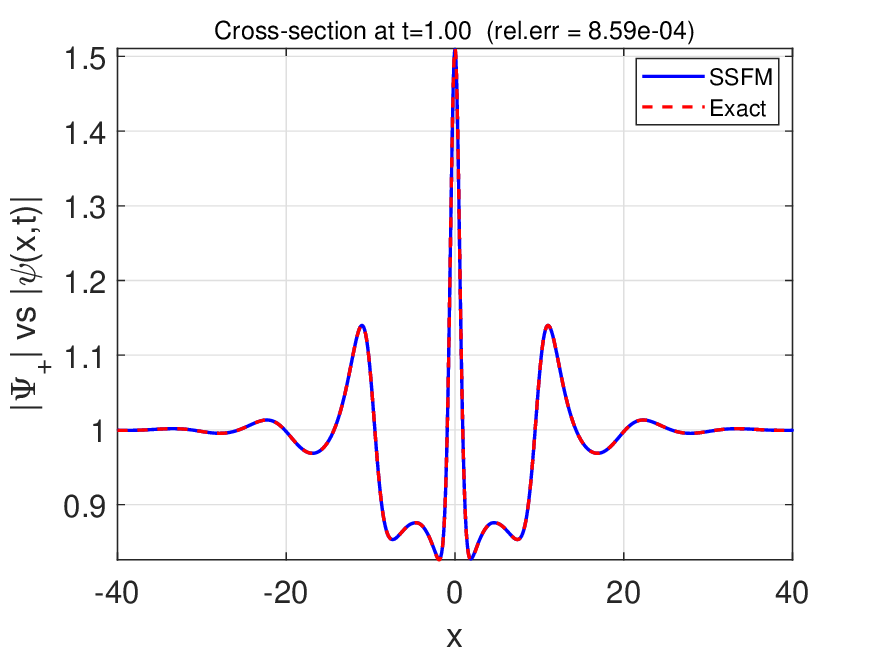}}}
{\rotatebox{0}{\includegraphics[width=4.8cm,height=3.8cm]{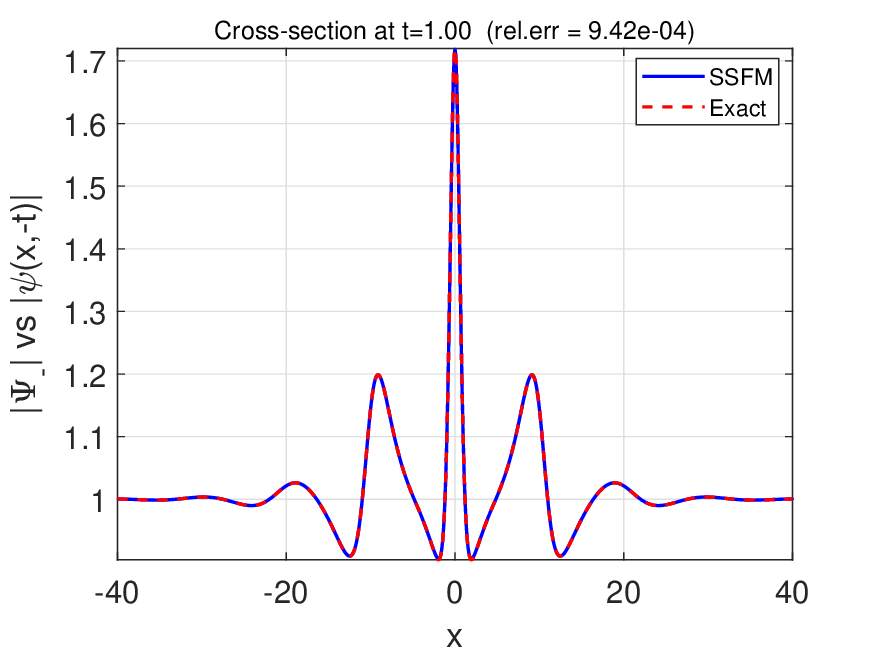}}}
{\rotatebox{0}{\includegraphics[width=4.8cm,height=3.8cm]{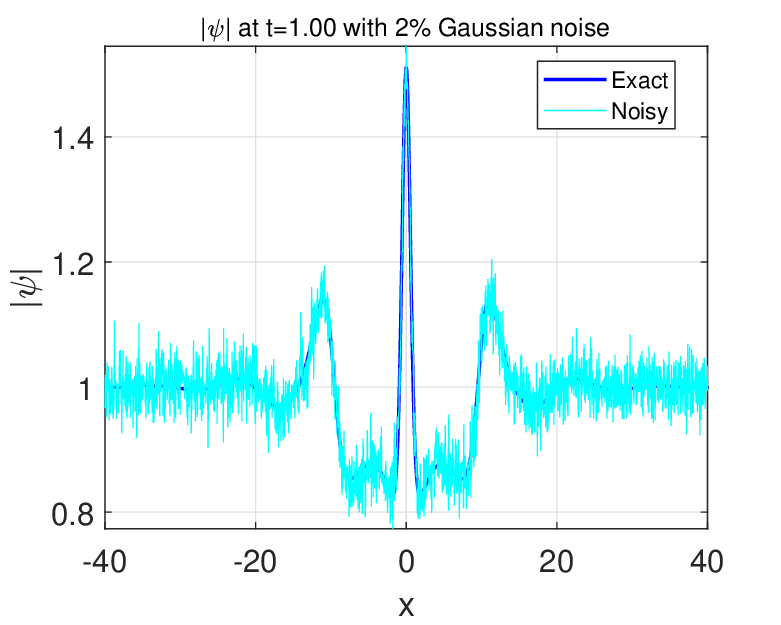}}}\\
$~~~~~~~~~~~~~~~~~~(\textbf{a})
~~~~~~~~~~~~~~~~~~~~~~~~~~~~~~~~~~~(\textbf{b})
~~~~~~~~~~~~~~~~~~~~~~~~~~~~~~~~~~~(\textbf{c})$\\

\noindent { \small \textbf{Figure 8.} Numerical simulations of the one-solitonic solution shown in Fig.\,1(a). $\textbf{(a)}$: Cross-section of numerical solution $|\Psi_+|$ and exact solution $|\psi|$ at $t=1$; $\textbf{(b)}$: Cross-section of numerical solution $\Psi_-$ and exact solution $|\psi(x,-t)|$ at $t=1$; $\textbf{(c)}$: Comparison of exact and noisy solutions at $t=1$ with $2\%$ additive noise.\\}

\section{Conclusion}\label{sec7}

This work systematically investigates superregular solitonic solutions of the reverse-time nonlocal NLSE in the defocusing regime. In contrast to the scalar defocusing NLSE, which does not admit superregular solitonic solutions, this work demonstrates the existence of superregular solitonic solutions for the reverse-time nonlocal NLSE in the defocusing regime. Based on the dressing method combined with the Joukowski transformation, the general analytical $N$-solitonic solution is constructed, and the explicit expressions for one-solitonic, two-solitonic, and superregular two-solitonic solutions are successfully derived. Further analysis reveals distinct multi-wave interaction structures in this system compared with the standard NLSE. Specifically, the one-solitonic solution is formed by the interaction of two wave components, the two-solitonic solution contains four wave components. Numerical simulations verify the effectiveness and robustness of the obtained analytical solutions, confirming the robustness of solitonic solutions under 2\% noise perturbations. These results enrich the theoretical framework of soliton dynamics in nonlocal nonlinear models and provide new physical insights for understanding wave propagation and MI mechanisms in nonlocal optical media.

\section*{Acknowledgements}
\hspace{0.3cm}
The authors are very grateful to Professor Evgeny Kuznetsov and Professor Shou-Fu Tian for their enthusiastic support and useful suggestions.
This work was supported by the Natural Science Foundation of China (Grant No. 12201622).

%\end{CJK*}
%\end{document}

% Converted from Microsoft Word to LaTeX
% by Chikrii SoftLab Word2TeX converter (version 2.4)
% Copyright (C) 1999-2001 Kirill A. Chikrii, Anna V. Chikrii
% Copyright (C) 1999-2001 Chikrii SoftLab.
% All rights reserved.
% http://www.word2tex.com/
% mailto: info@word2tex.com, support@word2tex.com

% Warning: You are using UNREGISTERED Chikrii SoftLab Word2TeX!
%          In UNREGISTERED mode some restrictions will apply.
%          For more information please visit http://www.word2tex.com/
% YOU CAN USE THIS FILE WITH THE SOLE PURPOSE OF EVALUATING Word2TeX.

%\documentclass [12pt]{article}

%\begin{document}

\end{document}